\documentclass{aa}  

\usepackage{graphicx}

\usepackage{amsmath,amssymb}
\usepackage[flushleft]{threeparttable}
\usepackage{array,booktabs,makecell}
\usepackage{enumitem}
\usepackage{multirow}
\usepackage[dvipsnames]{xcolor}

\usepackage{color}
\usepackage{xcolor}

\usepackage{txfonts}
\usepackage[]{hyperref}
\begin{document} 
   
   \title{FilDReaMS \\ 
   2. Application to the analysis of the relative orientations between filaments and the magnetic field in four \textit{Herschel} fields}
   
   \titlerunning{FilDReaMS. 2.
   Relative orientations of filaments to the magnetic field}

   \author{J.-S. Carrière
          \inst{1}
          \and
          K. Ferrière\inst{1}
          \and
          I. Ristorcelli\inst{1}
          \and
          L. Montier\inst{1}
          }

   \institute{IRAP, Université de Toulouse, CNRS, 9 avenue du Colonel Roche, BP 44346, 31028 Toulouse Cedex 4, France\\
              \email{jeansebastienpaulcarriere@gmail.com}
             }

   %\date{Received September 15, 1996; accepted March 16, 1997}

% \abstract{}{}{}{}{} 
% 5 {} token are mandatory
 
  \abstract
   {Both simulations and observations of the interstellar medium show that the study of the relative orientations between filamentary structures and the magnetic field can bring new insight into the role played by magnetic fields in the formation and evolution of filaments and in the process of star formation.}
   {We provide a first application of {\tt FilDReaMS}, the new method presented in the companion paper to detect and analyze filaments in a given image.
   The method relies on a template that has the shape of a rectangular bar with variable width. Our goal is to investigate the relative orientations between the detected filaments and the magnetic field.}
   {We apply {\tt FilDReaMS} to a small sample of four {\it Herschel} fields (G210, G300, G82, G202) characterized by different Galactic environments and different evolutionary stages. First, we look for the most prevalent bar widths, and we examine the networks formed by filaments of different bar widths as well as their hierarchical organization. Second, we compare the filament orientations to the magnetic field orientation inferred from {\it Planck} polarization data and, for the first time, we study the statistics of the relative orientation angle as functions of both spatial scale and ${\rm H_2}$ column density.}
   {We find preferential relative orientations in the four {\it Herschel} fields: small filaments with low column densities tend to be slightly more parallel than perpendicular to the magnetic field; in contrast, large filaments, which all have higher column densities, are oriented nearly  perpendicular (or, in the case of G202, more nearly parallel) to the magnetic field. In the two nearby fields (G210 and G300), we observe a transition from mostly parallel to mostly perpendicular relative orientations at an ${\rm H_2}$ column density $\simeq 1.1\times10^{21}\,$cm$^{-2}$ and $1.4\times10^{21}\,$cm$^{-2}$, respectively, consistent with the results of previous studies.}
   {Our results confirm the existence of a coupling between magnetic fields at cloud scales and filaments at smaller scale. They also illustrate the potential of combining {\it Herschel} and {\it Planck} observations, and they call for further statistical analyses with our dedicated method.}

   \keywords{ISM: clouds --
                ISM: structures --
                ISM: magnetic fields --
                dust --
                infrared: ISM --
                submillimiter: ISM --
                techniques: image processing
               }

   \maketitle
%
%-------------------------------------------------------------------

\section{Introduction}
\label{sec:introduction}

The physical processes involved in the early stages of star formation are still poorly understood. There is growing evidence that magnetic fields, together with turbulence and gravity, contribute significantly to the formation of stars, but their respective roles at different spatial and temporal scales are not clearly established \citep[][]{Mckee_2007, Dobbs_2014}. 
Making progress requires in-depth observational studies of the formation and evolution of dense structures over a broad range of scales, from molecular clouds down to filaments, clumps, and cores.

Filamentary structures were first detected in dust extinction \citep[][]{Schneider_1979}.
They were later observed in dust emission with the {\it Herschel} space observatory \citep[][]{Andre_2010, Molinari_2010, Menshchikov_2010, Miville-Deschenes_2010} and found to be ubiquitous in all Galactic (neutral) environments. 
Moreover, these observations showed that the many detected pre-stellar cores seemed to be mainly distributed in the densest filaments \citep[][]{Polychroni_2013, Konyves_2015, Montillaud_2015}. The question of the origin and evolution of dense cores is thus strongly connected to that of filaments \citep[see review by][]{Andre_2014}.

Filamentary structures are also observed to be dominant features in numerical simulations of magnetohydrodynamic (MHD) turbulence, indicating the potential influence of magnetic fields in their formation \citep[see review by][]{Hennebelle_2019}.
Furthermore, simulations make it clear that studying the relative orientation between magnetic fields and filaments can shed light on the role played by magnetic fields in the formation and evolution of filaments \citep[see for instance][]{Soler_2017, Wu_2017}. In particular, the simulations of  \citet[][]{Soler_2013} showed that in a weakly magnetized medium, the magnetic field is preferentially oriented parallel to the density structures, whereas in a strongly magnetized medium, there is a change in relative orientation from parallel to perpendicular at a critical density \citep[see also ][]{Chen_2016}.

A number of observational studies reveal preferential relative orientations between magnetic fields and filaments, as initially inferred from starlight polarization \citep[e.g.,][]{Goodman_1990, Pereyra_2004, Sugitani_2011, Palmeirim_2013, Li_2013, Clark_2014}. More recently, the statistical analysis based on the {\it Planck} survey of dust polarized emission showed that elongated structures are predominantly aligned parallel to the magnetic field in the diffuse (neutral) medium, while they are mostly perpendicular in dense molecular clouds \citep[][]{Bracco_2016, Soler_2016}, with a transition at a column density $N_{\rm H} \simeq 10^{21.7}\,$cm$^{-2}$ \citep[][]{Soler_2016}. 
A similar trend (low-$N_{\rm H}$ filaments mostly parallel and high-$N_{\rm H}$ filaments mostly perpendicular to the magnetic field) within individual clouds, with a similar transition column density, were also found
by \citet[][]{Malinen_2016, Cox_2016, Soler_2019_Herschel}, who compared the orientations of the plane-of-sky (PoS) magnetic field traced by {\it Planck} (at $\simeq$ 10' resolution) and the filaments traced by {\it Herschel} (36" resolution) in a sample of nearby molecular clouds.These results provide evidence that a coupling exists between magnetic fields at cloud scales and filaments at smaller scales \citep[][]{Soler_2019_Herschel}.
They also demonstrate the benefit of combining {\it Planck} and {\it Herschel} data sets despite their different angular resolutions.

This benefit, in turn, calls for further statistical analyses based on {\it Planck} and {\it Herschel} data together to explore relative orientation trends towards star-forming regions in different Galactic environments.
The {\it Herschel} Galactic cold core (GCC) key-program \citep[][]{Juvela_GCCI_2010, Juvela_GCCIII_2012}, which includes 116 Galactic fields ($\sim 40’\times 40’$ size on average) hosting {\it Planck} clumps ({\it Planck} catalogue of Galactic cold clumps (PGCC) from \citet[][]{Planck_XXVIII_2015}) is particularly well suited for such a statistical study. It covers a wide range of Galactic positions, environments, physical conditions, and evolutionary stages \citep[][]{Montillaud_2015}. Most of the fields display filaments over a broad range of column densities \citep[][]{Juvela_GCCIII_2012}, often including low-density striations (as in the GCC field L1642 studied by \citet[][]{Malinen_2016}). 
Future statistical analyses of relative orientations will require an efficient and robust method that makes it possible to extract filaments over a broad range of sizes and column densities and to determine their orientations.

The purpose of the companion paper (Carrière et al. 2022, hereafter Paper~1) was precisely to develop such a method, which we dubbed {\tt FilDReaMS} (Filament Detection \& Reconstruction at Multiple Scales).
Our purpose here is to apply {\tt FilDReaMS} to a first selection of four \textit{Herschel} fields and to study the properties of the detected filaments, with a special attention to their relative orientations to the magnetic field.

In Sect.~\ref{sec:data}, we present the data used for our study.
In Sect.~\ref{sec:FilDReaMS_nutshell}, we review our new {\tt FilDReaMS} method, giving just enough details for its application.
In Sect.~\ref{sec:analysis_results}, we apply {\tt FilDReaMS} to a sample of four {\it Herschel} fields selected amongst the 116 fields of the {\it Herschel}-GCC program.
In Sect.~\ref{sec:discussion}, we provide a detailed discussion of our results.
In Sect.~\ref{sec:conclusion}, we summarize and conclude our study.

\section{Data}
\label{sec:data}

We will apply our new {\tt FilDReaMS} method to a sample of four molecular clouds, for which we will exploit the combination of {\it Planck}-HFI and {\it Herschel} data.
In brief, {\it Planck}-HFI observations at 353\,GHz provide a whole-sky map of the dust polarized emission with an angular resolution of 4.7'; from this map, one can infer (a dust-weighted line-of-sight average of) the orientation of the PoS component of the magnetic field.
Meanwhile, {\it Herschel} observations provide dust emission maps at several wavelengths (with an angular resolution of 18" at $250\,{\rm \mu m}$); based on these maps, column density maps have been constructed with an angular resolution of 36".

\subsection{\textbf{\textit{Planck}}-HFI}
\label{ss:planck}

The polarized emission measured by {\it Planck}-HFI at 353\,GHz is dominated by the thermal emission from Galactic dust \citep{Planck_Int_XIX_2015}.
This emission, which is polarized perpendicular to the local magnetic field, provides a good probe of the magnetic field orientation in the densest regions of the interstellar medium (ISM).
The magnetic field PoS orientation angle, $\psi_B$, can be written in terms of the Stokes parameters, $Q$ and $U$, as

\begin{equation}
    \label{eq:Stokes}
    \psi_B = \frac{1}{2} \arctan{\left(\frac{U}{Q}\right)} \pm 90^\circ \ , \ ,
\end{equation}

\noindent where $\arctan$ is the two-argument arctangent function defined from $-180^\circ$ to $180^\circ$.
Since polarization data do not give access to the magnetic field direction, but only to its orientation, we may arbitrarily require that $\psi_B$ must lie in the range $[-90^{\circ}, +90^{\circ}]$; the $+$ or $-$ sign in the last term of Eq.~\ref{eq:Stokes} is then chosen accordingly.
Here, we adopt the IAU polarization convention in Galactic coordinates, such that $\psi_B$ increases counterclockwise from Galactic north. Since this is opposite to the {\tt Healpix} convention used in the {\it Planck} community, we have to change the sign of $U$ taken from the {\it Planck} Legacy Archive\footnote{{\it Planck} Legacy Archive: \url{https://www.cosmos.esa.int/web/planck/pla}}.

We extracted from the all-sky {\it Planck} $Q$ and $U$ maps at 353\,GHz a set of four much smaller maps ($2^{\circ}\times2^{\circ}$, approximately twice the size of {\it Herschel} maps), each centered on one of our four selected {\it Herschel} fields.
We smoothed the original {\it Planck} $Q$ and $U$ maps from 4.7' to 7' resolution.
The smoothing of these maps and their uncertainties was performed as described in Appendix of \citep{Planck_Int_XIX_2015}.
Our choice of a 7' resolution results from a compromise between the need to reach a high enough SNR (SNR > 3) in almost all pixels and our wish to access the smallest possible magnetic field fluctuations.
The typical improvement of polarization SNR when smoothing from 4.7' to 7' is roughly a factor of $7' / 4.7' \simeq 1.5$;
in the case of our four fields, the fraction of pixels with SNR < 3 in the {\it Planck} $Q$ and $U$ maps smoothed to 7' is $\lesssim 1\,\%$.

The uncertainty in the polarization angle was computed as described in the appendix of \cite{Planck_Int_XIX_2015}.
Following \cite{Malinen_2016}, we only retained pixels with uncertainty $< 10^\circ$.
It is difficult to directly estimate the impact of our smoothing on the inferred polarization angles, because the SNR at 4.7' resolution is not high enough.
However, we verified that smoothing from 7' to 10' (two resolutions leading to SNR > 3 in almost all pixels) had very little impact on the derived polarization angles.
We are, therefore, confident that our results will not be significantly affected by our choice of resolution for the {\it Planck} $Q$ and $U$ maps.

\subsection{\textbf{\textit{Herschel}}}

\subsubsection{{\it Herschel} data}
\label{ss:herschel_PGCC}

The {\it Herschel} satellite \citep{Pilbratt_2010} provided a fantastic probe of filaments within molecular clouds at a much better angular resolution than that of {\it Planck}. The {\it Herschel} open-time key program Galactic Cold Cores \citep{Juvela_GCCI_2010}, which we will refer to as the {\it Herschel}-GCC program, was designed to map out a sample of cold regions of interstellar clouds previously detected in the {\it Planck} all-sky survey.
This follow-up is composed of 116 fields (with a typical size of 40') observed with PACS and SPIRE \citep[][]{Poglitsch_2010,Griffin_2010}, which were selected using the Planck Catalogue of Galactic cold clumps \citep[PGCC,][]{Planck_XXVIII_2015} to cover a diversity of clump physical properties and Galactic environments \citep{Juvela_GCCIII_2012}. The fields are presented in detail in  \citet{Montillaud_2015}, who built a catalogue of $\approx 4000$ cold and compact sources, identifying their various evolutionary stages from gravitationally unbound to prestellar and protostellar cores. 

In this study, we use the H$_2$ column density maps presented in \citet{Montillaud_2015} and derived from the dust spectral energy distribution (SED) fit based on the $250$, $350$, and $500\,{\rm \mu m}$ SPIRE maps. The authors assumed a fixed emissivity spectral index $\beta = 2.0$, with a dust opacity $\kappa = 0.1\,{\rm cm^{-2}/g}\,(\nu/1000\,{\rm GHz})^{\beta}$. The $N_{\rm H_2}$ maps have an angular resolution of 36".

\subsubsection{Sample of four fields from the {\it Herschel}-GCC program}

\begin{figure*}
    \centering
    \includegraphics[width=\textwidth]{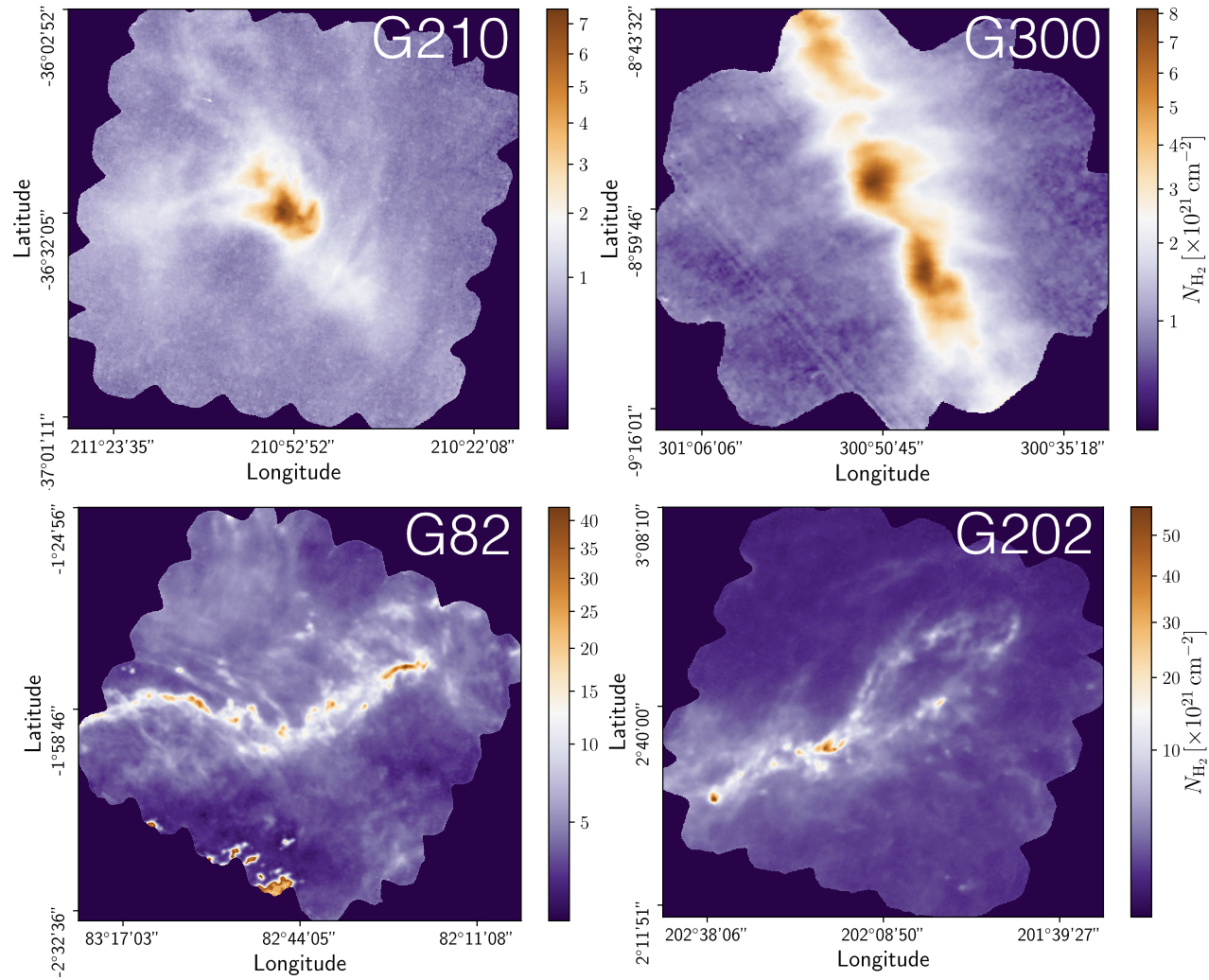}
    \caption{${\rm H_2}$ column density maps of the four {\it Herschel} fields selected from the {\it Herschel}-GCC program (see Sect.~\ref{ss:herschel_PGCC}).}
    \label{fig:NH_fields}
\end{figure*}

We focus on four {\it Herschel} fields out of the 116 fields of the {\it Herschel}-GCC program.
Their characteristics are summarized in Table~\ref{tab:input_fields}, while their H$_2$ column density maps are displayed in Fig.~\ref{fig:NH_fields}.
The reason why we selected these four molecular clouds is because they present a good diversity, which allows us to explore various Galactic environments and filament morphologies. Furthermore, they are already well-known and benefit from both observations with multiple tracers and various in-depth analyses carried out in the past decade, which provided us with information concerning their evolutionary stage or their dynamics (see references below).

The G210 molecular cloud is located at high Galactic latitude ($\vert b \vert \simeq 36^{\circ}$) in a diffuse environment. It was studied by \citet[][]{Malinen_2014, Malinen_2016}, who found an unusually high star formation rate (SFR) considering its low mass, as well as a complex interplay between the large-scale magnetic field and the cloud structure.

The G300 cloud lies at medium Galactic latitude ($\vert b \vert \simeq 9^{\circ}$) inside the well-known Musca filament. It was part of a statistical analysis by \citet[][]{Arzoumanian_2019}, and it was studied in greater detail by \citet[][]{Cox_2016} and \citet[][]{Kainulainen_2016}. It was found to be a filament at a very early stage of evolution with an SFR $\simeq 1\%$ \citep[][]{Cox_2016}, and it was shown to be mainly thermally super-critical and fragmenting faster than the center of the Musca filament \citep[fragments 4 and 5 in][]{Kainulainen_2016}.

The G82 cloud is located close to the Galactic plane ($\vert b \vert \simeq 2^{\circ}$).
It was found by \citet[][]{Saajasto_2017} to be a filament at a late stage of evolution, and it was described as a "debris filament" already forming several cold cores and young stellar objects.

Finally the G202 cloud, also close to the Galactic plane ($\vert b \vert \simeq 3^{\circ}$), is part of the Monoceros B molecular complex. It was studied in great detail by \citet[][]{Montillaud_2019_1, Alina_2021}. 
They both found this filament to be thermally super-critical. \citet{Montillaud_2019_1, Montillaud_2019_2} also found that this filament is actively forming stars, especially in its center, the "junction" region, which lies between two colliding filaments and represents $\simeq 60\%$ of the cloud mass with a high SFR $\simeq 20-38\%$.

\begin{table}
\caption{Basic parameters of the ${\rm H_2}$ column density maps of the four {\it Herschel} fields displayed in Fig.~\ref{fig:NH_fields}.}
\begin{threeparttable}
\begin{tabular}{m{1.7cm} m{6.4cm}}
\toprule\toprule
\textbf{\large G210}
& $l = 210.90^{\circ}$\\
& $b = -36.55^{\circ}$\\
& $d = 140\pm20\,{\rm pc}$ \\
& $1\,{\rm px} = 12" \rightarrow 0.0081\,{\rm pc}$ \\
& $\Delta l \times \Delta b = 1.28^{\circ}\times1.22^{\circ} \rightarrow 3.1\,{\rm pc}\times3.0\,{\rm pc}$\\
& $N_{\rm H_2} \in [0.2, 7.5]\times10^{21}\,{\rm cm}^{-2}$\\ 

\midrule
\textbf{\large G300}
& $l = 300.86^{\circ}$\\
& $b = -9.00^{\circ}$\\
& $d = 150\pm30\,{\rm pc}$ \\
& $1\,{\rm px} = 12" \rightarrow 0.0087\,{\rm pc}$ \\
& $\Delta l \times \Delta b = 0.64^{\circ}\times0.68^{\circ} \rightarrow 1.7\,{\rm pc}\times1.8\,{\rm pc}$\\
& $N_{\rm H_2} \in [0.6, 8.2]\times10^{21}\,{\rm cm}^{-2}$\\ 

\midrule
\textbf{\large G82}
& $l = 82.65^{\circ}$\\
& $b=-2.00^{\circ}$\\
& $d = 620^{+31}_{-42}\,{\rm pc}$ \\
& $1\,{\rm px} = 12" \rightarrow 0.036\,{\rm pc}$ \\
& $\Delta l \times \Delta b = 1.37^{\circ}\times1.41^{\circ} \rightarrow 14.9\,{\rm pc}\times15.3\,{\rm pc}$\\
& $N_{\rm H_2} \in [2.7, 42]\times10^{21}\,{\rm cm}^{-2}$\\ 
\midrule
\textbf{\large G202}
& $l = 202.02^{\circ}$\\
& $b = +2.85^{\circ}$\\
& $d = 760\pm100\,{\rm pc}$ \\
& $1\,{\rm px} = 12" \rightarrow 0.044\,{\rm pc}$ \\
& $\Delta l \times \Delta b = 1.22^{\circ}\times1.17^{\circ} \rightarrow 16.2\,{\rm pc}\times15.6\,{\rm pc}$\\
& $N_{\rm H_2} \in [0.6, 58]\times10^{21}\,{\rm cm}^{-2}$\\ 
\bottomrule\bottomrule
\end{tabular}
  \begin{tablenotes}
    \item {\bf Notes.} Parameters are ordered by increasing distance: Galactic longitude, $l$, Galactic latitude, $b$, distance $d$ (taken from \citet{Saajasto_2017} for G82 and \citet{Montillaud_2015} for the other fields), angular and physical sizes of a pixel  (equal to one-third of the beam size), angular and physical dimensions of the map, $\Delta l \times \Delta b$, and range of the ${\rm H_2}$ column density, $N_{\rm H_2}$.
  \end{tablenotes}
\end{threeparttable}
\label{tab:input_fields}
\end{table}

\section{\texttt{FilDReaMS} in a nutshell}
\label{sec:FilDReaMS_nutshell}

\subsection{Overview of the method}
\label{sec:FilDReaMS_overview}

The purpose of {\tt FilDReaMS} (described in details in Paper 1) is to detect and characterize filamentary structures in an image, which in the present paper is a {\it Herschel} (intensity or column-density) map.
To that end, {\tt FilDReaMS} resorts to a special template, which has the shape of a rectangular bar (referred to as the model bar) of width $W_{\rm b}$, length $L_{\rm b}$, and aspect ratio $r_{\rm b} = L_{\rm b} / W_{\rm b}$.
In the following astrophysical applications, we adopt $r_{\rm b} = 3$ \citep{Panopoulou_2014, Arzoumanian_2019} and we consider values of $W_{\rm b}$ spanning the range $[(W_{\rm b})_{\rm min}, (W_{\rm b})_{\rm max}]$, with $(W_{\rm b})_{\rm min} = 5\,{\rm px}$ and $(W_{\rm b})_{\rm max}$ equal to one-ninth the size of the {\it Herschel} map.
The orientation angle of the model bar, $\psi_{\rm b}$, follows the same convention as the magnetic field orientation angle, $\psi_B$ (Eq.~\ref{eq:Stokes}), such that $\psi_{\rm b}$ is defined in the range $[-90^{\circ}, +90^{\circ}]$ and increases counterclockwise from Galactic north.
The same will hold true for the orientation angle of a filament, $\psi_{\rm f}$, defined in the next paragraph.

For any given value of $W_{\rm b}$, {\tt FilDReaMS} filters out structures broader than $W_{\rm b}$ in the initial image and converts the filtered image into a binary map.
At each pixel $i$ of the binary map, {\tt FilDReaMS} considers a model bar centered on $i$, derives the bar orientation angle that provides the best match to the binary map, $(\psi_{\rm b})_i$,
and computes the corresponding significance, $S_i$, based on a comparison with an ideal case.
If $S_i>1$, {\tt FilDReaMS} concludes that a significant filament with orientation angle $(\psi_{\rm f})_i = (\psi_{\rm b})_i$ is detected at pixel $i$.
The true shape of this filament is then reconstructed from the binary map and its intensity from the initial image.
Iterating over all the pixels of the binary map (minus a band of width $L_{\rm b}/2$ adjacent to the border) and superposing all the associated reconstructed filaments yields the entire network of physical filaments of bar width $W_{\rm b}$.
Each pixel $i'$ of this network is assigned a filament orientation angle, $(\psi_{\rm f}^\star)_{i'}$, equal to the orientation angle of the most significant filament (filament with the highest $S_i$) whose model bar contains $i'$.

The procedure is repeated for different values of $W_{\rm b}$, each of which leads to a new network of filaments.
Each pixel $i'$ of the different networks is assigned a most significant bar width, $(W_{\rm b}^\star)_{i'}$, equal to the bar width of the most significant filament whose model bar contains $i'$.
Finally, the histogram of $W_{\rm b}^\star$ over all pixels makes it possible to identify the most prevalent bar widths for the entire map, $W_{\rm b}^{\star{\rm peak}}$.
In the following, to make it easier to compare the different {\it Herschel} fields, we will work with the normalized histogram $(N_{\rm pix}/N_{\rm map})$ versus $W_{\rm b}^\star$, where $N_{\rm pix}$ is the number of pixels whose most significant bar width is $W_{\rm b}^\star$ and $N_{\rm map}$ is the total number of pixels in the map.

For convenience, the meaning of all the parameters introduced above can be found in Table~\ref{tab:FilDReaMS_notations}.

\begin{table*}
\caption{List of all the symbols used in the paper.}
\centering
\begin{threeparttable}
\begin{tabular}{m{2.5cm} m{10cm}}
\midrule\midrule
$i$ & Index of the considered pixel in the binary  map\\
\cmidrule(l  r ){1-2}
$i'$ & Index of the considered pixel in the map of reconstructed filaments\\
\cmidrule(l  r ){1-2}
$\psi_{\rm B}$ & Orientation angle of the PoS component of the magnetic field\\
\cmidrule(l  r ){1-2}
$W_{\rm b}$ & Width of the model bar\\
\cmidrule(l  r ){1-2}
$L_{\rm b}$ & Length of the model bar\\
\cmidrule(l  r ){1-2}
$r_{\rm b} = L_{\rm b}/W_{\rm b}$ & Aspect ratio of the model bar\\
\cmidrule(l  r ){1-2}
$\psi_{\rm b}$ & Orientation angle of the model bar\\
\cmidrule(l  r ){1-2}
$S$ & Significance of a filament detection\\
\cmidrule(l  r ){1-2}
$\psi_{\rm f}$ & Orientation angle of a filament\\
\cmidrule(l  r ){1-2}
$\psi_{\rm f}^{\star}$ & Orientation angle of the most significant filament\\
\cmidrule(l  r ){1-2}
$W_{\rm b}^{\star}$ & bar width of the most significant filament\\
\cmidrule(l  r ){1-2}
$N_{\rm pix}$ & Number of pixels whose most significant bar width is $W_{\rm b}^\star$\\
\cmidrule(l  r ){1-2}
$N_{\rm map}$ & Total number of pixels in the map\\
\cmidrule(l  r ){1-2}
$W_{\rm b}^{\star{\rm peak}}$ & Most prevalent bar width for the entire map\\
\midrule\midrule
\end{tabular}
\end{threeparttable}
\label{tab:FilDReaMS_notations}
\end{table*}

\subsection{Applications}
\label{sec:application_orientation}

A first obvious application of {\tt FilDReaMS} concerns the bar widths of filaments, with, in particular, the idea of uncovering the most prevalent bar widths.
Related goals are the visualization of the network formed by filaments of a given bar width and the examination of the spatial connections between filaments of different bar widths.
All these, in turn, may provide some insight into the process of filament formation.

A second, very important application of {\tt FilDReaMS}, which we will devote most of our attention to, concerns the orientations of filaments relative to the local magnetic field. 
The idea is to compare the filament orientation angles, $\psi_{\rm f}^{\star}$, derived with {\tt FilDReaMS} to the magnetic field PoS orientation angle, $\psi_B$, inferred from the dust polarized emission observed by {\it Planck} (Eq.~\ref{eq:Stokes}).
The natural quantity to work with in this context is the relative orientation angle, $(\psi_{\rm f}^{\star} - \psi_B)$.
For convenience, since both $\psi_{\rm f}^{\star}$ and $\psi_B$ are defined in the range $[-90^{\circ}, +90^{\circ}]$ (see Sects.~\ref{sec:FilDReaMS_overview} and \ref{ss:planck}, respectively), we require that $(\psi_{\rm f}^{\star} - \psi_B)$ must also lie in the range $[-90^{\circ}, +90^{\circ}]$;
if needed, this can be achieved by adding or subtracting $180^{\circ}$.
{\tt FilDReaMS} will enable us to study, for the first time, the statistics of $(\psi_{\rm f}^{\star} - \psi_B)$ as functions of spatial scale (Sect.~\ref{sec:filament_relative_orientation_scale_dependence}) and as functions of both spatial scale and ${\rm H_2}$ column density (Sect.~\ref{sec:filament_relative_orientation_variation_NH2}).

Following previous studies \citep{Soler_2013, Bracco_2016, Soler_2016, Soler_2019_Herschel}, we will construct histograms of relative orientation (HROs), which are just histograms of $(\psi_{\rm f}^{\star} - \psi_B)$, similar to histograms $(N_{\rm pix}/N_{\rm map})$ versus $(\psi_{\rm f}^{\star} - \psi_B)$.
The novelty is that we will do so in three separate ranges of bar widths.
We will also construct 2D histograms of $(\psi_{\rm f}^{\star} - \psi_B)$ as functions of ${\rm H_2}$ column density, $N_{\rm H_2}$ (i.e., 2D histograms $(N_{\rm pix}/N_{\rm map})$ versus $(N_{\rm H_2}, (\psi_{\rm f}^{\star} - \psi_B))$) 
again in three separate ranges of bar widths.
This will enable us to explore the idea of a bi-modal distribution of $(\psi_{\rm f}^{\star} - \psi_B)$, with a transition at a certain $N_{\rm H_2}$.
To be more quantitative, we will resort to the multivariate analysis methods presented by \citet{Malinen_2016},
namely, a non-negative matrix factorization \citep[NMF,][]{Lee_1999} and a principal component analysis \citep[PCA,][]{Joliffe_2002}. 
NMF makes it possible to derive the first (in our case, the first two) principal components of the 2D histograms, together with their respective weights (i.e., their relative contributions) in each $N_{\rm H_2}$ bin.
PCA estimates the capability to reconstruct the entire 2D histograms by including only the considered (in our case, the first two) principal components. This reconstruction capability quantifies the relevance of a bi-modal relative orientation.

\begin{figure*}
    \centering
    \includegraphics[width=\textwidth]{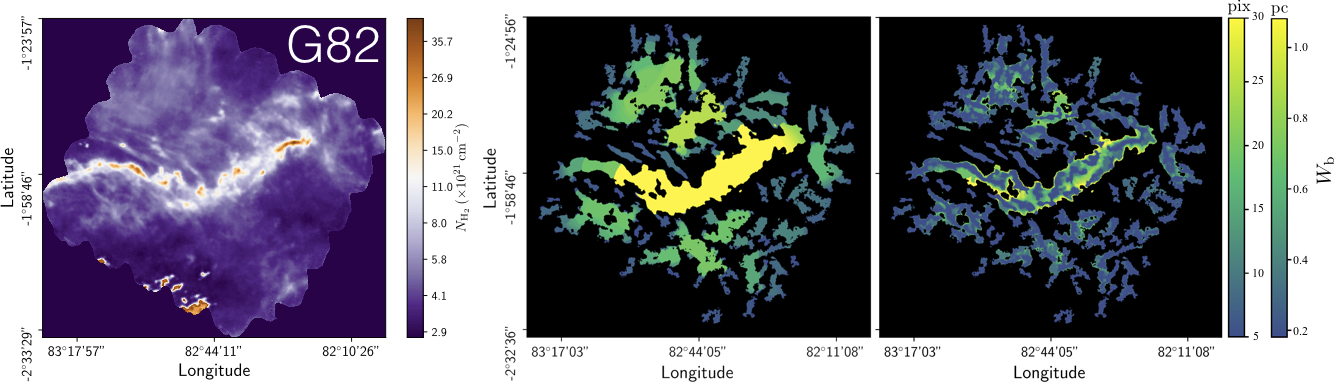}
    \caption{{\it Herschel} G82 field and all the filaments obtained with {\tt FilDReaMS}. 
    The size of the image is $414~{\rm px} \times 425~{\rm px}$. {\bf Left}: ${\rm H_2}$ column density map of G82. {\bf Middle and right}: Reconstructed filaments, with their bar widths, $W_{\rm b}$, in color (in pixels and in parsecs). For every pixel that belongs to filaments with different $W_{\rm b}$, the largest ({\bf Middle}) or the smallest ({\bf Right}) $W_{\rm b}$ is shown in the foreground.}
    \label{fig:filaments_G82}
\end{figure*}

\section{Results}
\label{sec:analysis_results}

We now present and analyze the results obtained with {\tt FilDReaMS} for a sample of four {\it Herschel} fields. We focus on two important characteristics of the reconstructed filaments: their bar widths, $W_{\rm b}$ (Sect.~\ref{sec:filament_barwidths}), and their orientations relative to the magnetic field, with a detailed discussion of how the relative orientation angles, $(\psi_{\rm f}^{\star} - \psi_B)$, statistically vary with bar width (Sect.~\ref{sec:filament_relative_orientation_scale_dependence}) and with column density (Sect.~\ref{sec:filament_relative_orientation_variation_NH2}).
In each subsection, we start with the G82 field, which exhibits a rich filamentary network and covers the broadest range of scales; we continue with the G210 field (only in Sects.~\ref{sec:filament_relative_orientation_scale_dependence} and \ref{sec:filament_relative_orientation_variation_NH2}), which shows interesting new features; and we finish with the general trends emerging from our small sample.
Detailed plots of the results obtained for the four individual {\it Herschel} fields are provided in Appendix~\ref{sec:summary_results_coldensity},
and a summary of the main characteristics of the reconstructed filaments can be found in Table~\ref{tab:sum_fields}.

\subsection{Filament bar widths}
\label{sec:filament_barwidths}

\subsubsection{G82}
\label{sec:filament_barwidths_G82}

Following the guidelines given in Sect.~\ref{sec:FilDReaMS_overview}, we restrict the values of $W_{\rm b}$ to the range $\left[5, 30\right]\,{\rm px}$. The corresponding range in parsecs can be derived using the relation

\begin{equation}
    \label{eq:physical_Wb}
    W_{\rm b}^{\left[\rm pc\right]} = 
    d \ \gamma \ W_{\rm b}^{\left[\rm px\right]}\,,
\end{equation}

\noindent where $d$ is the distance of the {\it Herschel} field (in pc) and $\gamma$ the angular size of a pixel (in rad). With $d=620^{+31}_{-42}\,$pc and $\gamma = 12''$ (see Table~\ref{tab:input_fields}), the range in parsecs is $\left[0.18, 1.08\right]\,$pc. The uncertainty in $W_{\rm b}^{\left[\rm pc\right]}$, $\Delta W_{\rm b}^{\left[\rm pc\right]}$, is related to the uncertainty in $W_{\rm b}^{\left[\rm px\right]}$, $\Delta W_{\rm b}^{\left[\rm px\right]} = 1\,{\rm px}$, and the uncertainty in $d$, $\Delta d$, through

\begin{equation}
    \label{eq:physical_deltaWb}
    \Delta W_{\rm b}^{\left[\rm pc\right]} = W_{\rm b}^{\left[\rm pc\right]}\left(\frac{\Delta d}{d} + \frac{\Delta W_{\rm b}^{\left[\rm px\right]}}{W_{\rm b}^{\left[\rm px\right]}}\right)\,.
\end{equation}

\noindent Hence, with $\Delta d / d \simeq 6\%$ and $\Delta W_{\rm b}^{\left[\rm px\right]}/W_{\rm b}^{\left[\rm px\right]}$ ranging from $1\,{\rm px} / 30\,{\rm px} \simeq 3\%$ to $1\,{\rm px} / 5\,{\rm px} = 20\%$, we find that $\Delta W_{\rm b}^{\left[\rm pc\right]}/W_{\rm b}^{\left[\rm pc\right]}$ ranges from $\simeq$ 3$\%$ to 19$\%$. From now on, for simplicity, we give the values of $W_{\rm b}^{\left[\rm pc\right]}$ without their uncertainties.

{\tt FilDReaMS} allows us to detect a wide variety of filaments, mostly visible with the naked eye in the initial {\it Herschel} G82 map (left panel of Fig.~\ref{fig:filaments_G82}). Filaments are detected in all areas of the map, over a broad range of sizes and column densities, with $N_{\rm H_2}$ varying from $\simeq 3\times10^{21}\,$cm$^{-2}$ to $\simeq 4\times10^{22}\,$cm$^{-2}$. 

All the reconstructed filaments are plotted in the middle and right panels of Fig.~\ref{fig:filaments_G82}. Because many pixels belong to several filaments (usually filaments of different bar widths), it is not possible to vizualize all the filaments in a single plot. Therefore, we show two extreme views, in which filaments of different bar widths are superposed, either with the smallest filaments in the background and the largest in the foreground (middle panel) or vice-versa (right panel). A fraction of the smaller filaments appear to be spatially connected to larger filaments, forming either crests, internal sub-structures (such as strands), or ramifications (which appear to emerge from, or merge with, the larger filaments).
The rest of the smaller filaments are disconnected from larger filaments.
Both ramifications and disconnected filaments can form striation patterns, faint and periodic structures similar to those detected by \citet[][]{Goldsmith_2008} and \citet[][]{Narayanan_2008} in the diffuse $^{12}$CO emission map from the Taurus molecular cloud.

Crests are roughly parallel to their parent filaments, while strands are observed at all angles.
Ramifications tend to connect with their parent filaments at large (roughly perpendicular) angles.
Disconnected small filaments are found at all angles, although in the vicinity of a large filament they tend to line up roughly parallel or perpendicular to this large filament.

\begin{figure}
    \centering
    \includegraphics[width=\columnwidth]{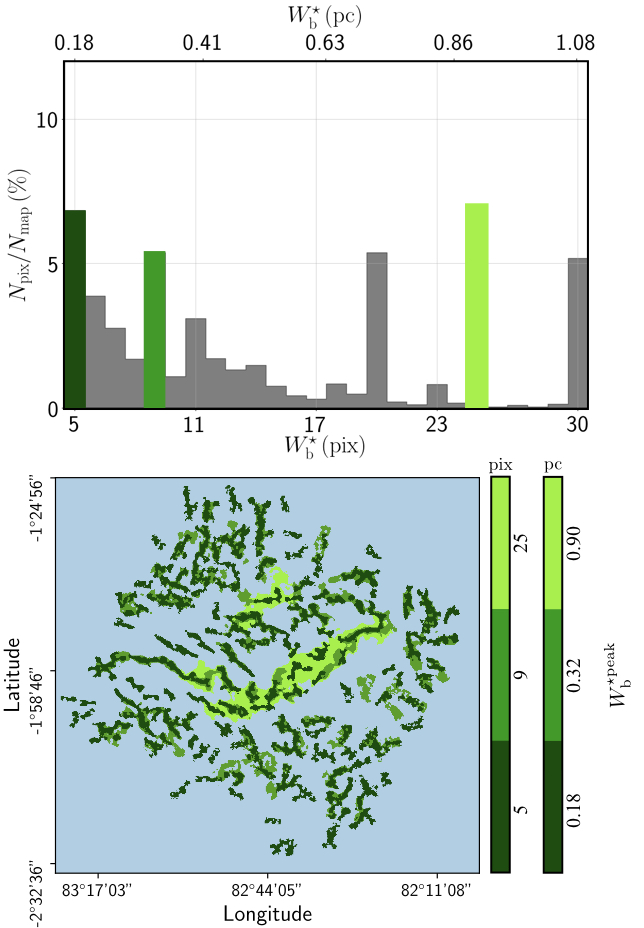}
    \caption{Results obtained for the bar widths in the {\it Herschel} G82 field.
    {\bf Top}: Number of pixels, $N_{\rm pix}$, whose most significant bar width is $W_{\rm b}^{\star}$, normalized to the number of pixels in the map, $N_{\rm map}$, as a function of $W_{\rm b}^{\star}$. The peaks of the histogram correspond to the most prevalent bar widths, $W_{\rm b}^{\star{\rm peak}}$, three of which (highlighted in green) were selected for visualization.
    {\bf Bottom}: Reconstructed filaments with bar width equal to one of the three selected $W_{\rm b}^{\star{\rm peak}}$.
    Smaller filaments are overlaid on top of larger filaments.}
    \label{fig:filament_best_sizes_G82}
\end{figure}

The histogram $(N_{\rm pix}/N_{\rm map})$ versus $W_{\rm b}^\star$ (defined in Sect.~\ref{sec:FilDReaMS_overview}) is displayed in the top panel of Fig.~\ref{fig:filament_best_sizes_G82}. Five peaks clearly emerge at $W_{\rm b}^{\star{\rm peak}}=5\,{\rm px}$, $9\,{\rm px}$, $20\,{\rm px}$, $25\,{\rm px}$, and $30\,{\rm px}$, corresponding to $0.18\,$pc, $0.32\,$pc, $0.72\,$pc, $0.90\,$pc, and $1.08\,$pc, respectively. These peaks represent the most prevalent bar widths. The peak at $W_{\rm b}^{\star{\rm peak}}=5\,{\rm px}$ is probably partly artificial, because it arises at the lower boundary of the $W_{\rm b}^\star$ range and, therefore, includes not only the contribution from filaments with $W_{\rm b}^\star=5\,{\rm px}$, but also the contribution from all the enclosed smaller filaments (not considered in this study). However, the fact that $N_{\rm pix}/N_{\rm map}$ rises almost continuously from $W_{\rm b}^\star = 8\,{\rm px}$ all the way in to (at least) $W_{\rm b}^\star = 5\,{\rm px}$ speaks in favor of a true physical peak at  $W_{\rm b}^{\star{\rm peak}} \le 5\,{\rm px}$. 
The peak at $30\,{\rm px}$, which arises at the upper boundary of the $W_{\rm b}^\star$ range, might also be partly artificial, in the sense that it could correspond to the crest of a larger filament with higher significance $S$.
In any case, the peaks at $9\,{\rm px}$, $20\,{\rm px}$ and $25\,{\rm px}$ are probably real.

In the bottom panel of Fig.~\ref{fig:filament_best_sizes_G82}, we show three sets of reconstructed filaments, each with a bar width $W_{\rm b}$ equal to one of the most prevalent bar widths. We choose $W_{\rm b}^{\star{\rm peak}}=5\,{\rm px}$ (dark green), $9\,{\rm px}$ (middle green), and $25\,{\rm px}$ (light green). In case of overlap, we let the smaller filaments appear in the foreground.
Together, the three sets of filaments provide a good picture of the entire filamentary network displayed in Fig.~\ref{fig:filaments_G82}, while they bring out the different morphological trends more clearly.

\subsubsection{The four {\it Herschel} fields}
\label{sec:filament_barwidths_all_fields}

\begin{figure}
    \centering
    \includegraphics[width=8.cm]{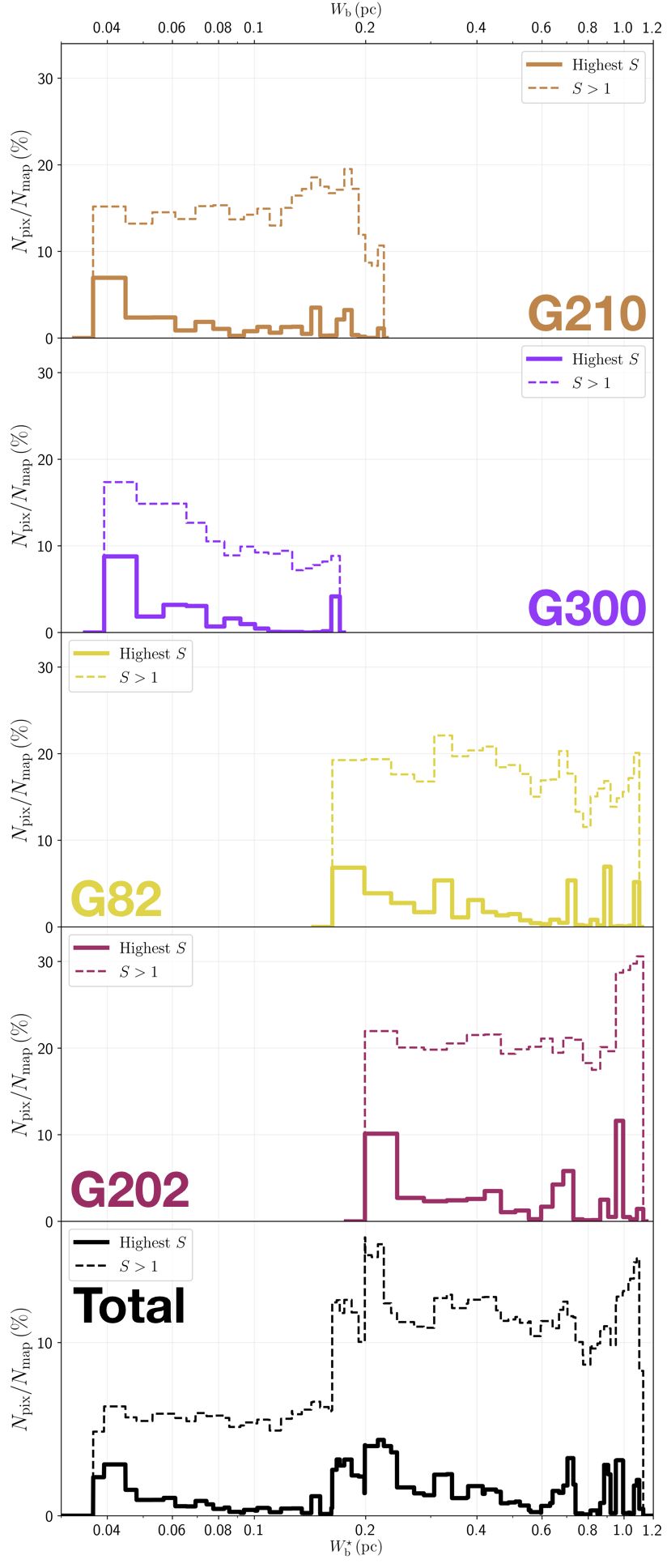}
    \caption{Results obtained for the bar widths in the four {\it Herschel} fields.
    {\bf Top four panels:} Histograms of bar width in the four individual {\it Herschel} fields.
    {\bf Bottom panel:} Combined histograms of the four {\it Herschel} fields together. In each panel, the solid line gives the normalized number of pixels whose most significant bar width is $W_{\rm b}^{\star}$ as a function of $W_{\rm b}^{\star}$, whereas the dashed line gives the normalized number of pixels belonging to a reconstructed filament of bar width $W_{\rm b}$ as a function of $W_{\rm b}$.}
    \label{fig:average_significance_fields}
\end{figure}

\begin{figure}
    \centering
    \includegraphics[width=\columnwidth]{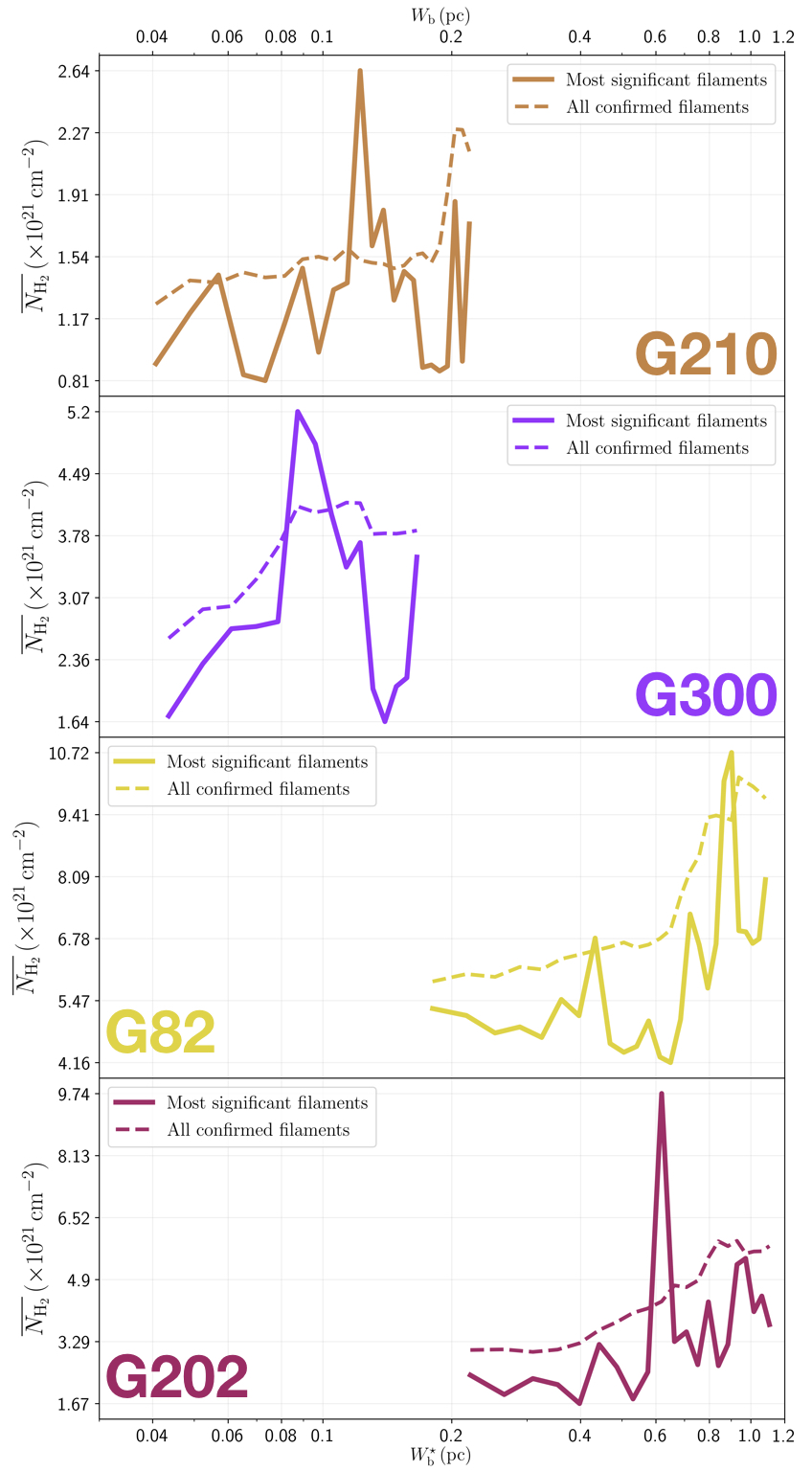}
    \caption{Average column density, $\overline{N_{\rm H_2}}$, as a function of bar width, $W_{\rm b}^{\star}$ (solid line) or $W_{\rm b}$ (dashed line), in each of the four {\it Herschel} fields. In each panel, the solid line gives the average over all the pixels whose most significant bar width is $W_{\rm b}^\star$, whereas the dashed line gives the average over all the pixels belonging to a reconstructed filament of bar width $W_{\rm b}$.}
    \label{fig:NH2_vs_Wb}
\end{figure}

The counterparts of Figs.~\ref{fig:filaments_G82} and \ref{fig:filament_best_sizes_G82} for the four {\it Herschel} fields are plotted in separate sheets in Appendix~\ref{sec:summary_results_coldensity}.
The interconnection between filaments of different bar widths observed in G82 is also seen in the other {\it Herschel} fields.
The reconstructed maps of G202 (Fig.~\ref{fig:G202_plots}) look similar to those of G82 (Fig.~\ref{fig:G82_plots}), with this small difference that G202 contains no visible striations.
In G210 (Fig.~\ref{fig:G210_plots}), a few striations are again present, but the largest filaments do not show the same degree of internal sub-structuring.
All this is even more true in G300 (Fig.~\ref{fig:G300_plots}), where, in addition to the more numerous striations, crests are visible inside the largest filaments.
More globally, G300 shows a higher degree of large-scale ordering, with most filaments being roughly aligned along two orthogonal directions.

The histograms $(N_{\rm pix}/N_{\rm map})$ versus $W_{\rm b}^\star$ of the four individual {\it Herschel} fields are displayed in Fig.~\ref{fig:average_significance_fields} (top four panels, solid line), along with the combined histogram of the four {\it Herschel} fields together (bottom panel, solid line).
These histograms include only the most significant filaments, which means that for every value of $W_{\rm b}^\star$, they give the normalized number of pixels whose most significant bar width is $W_{\rm b}^\star$.
For completeness, we also plot the histograms $(N_{\rm pix}/N_{\rm map})$ versus $W_{\rm b}$ that include all the reconstructed filaments (dashed line), which means that for every value of $W_{\rm b}$, they give the normalized number of pixels belonging to a reconstructed filament of bar width $W_{\rm b}$. 
All the histograms are plotted along a common $W_{\rm b}^\star$ (or $W_{\rm b}$) axis, which encompasses the different $W_{\rm b}^\star$ ranges of the individual {\it Herschel} fields ($\simeq [0.04, 0.2]\,{\rm pc}$ for G210 and G300 and $\simeq [0.2, 1.1]\,{\rm pc}$ for G82 and G202).

Each of the individual histograms $(N_{\rm pix}/N_{\rm map})$ versus $W_{\rm b}^\star$ (solid line) contains between 4 and 6 peaks.
As explained for G82 (Sect.~\ref{sec:filament_barwidths_G82}), the first and last peaks, which arise at the lower and upper boundaries, could be partly artificial.
Accordingly, the peaks near $0.04\,{\rm pc}$, $0.2\,{\rm pc}$, and $1.1\,{\rm pc}$ in the combined histogram could also be partly artificial.
The other peaks in the histograms of G210 and G300 are probably too weak to be truly significant.
In contrast, both G82 and G202 exhibit two strong peaks near $0.7\,{\rm pc}$ and $0.9\,{\rm pc}$, which clearly stand out in the combined histogram.
No such peaks appear in the histograms $(N_{\rm pix}/N_{\rm map})$ versus $W_{\rm b}$ (dashed line), where the contributions from the most significant filaments are lost in the contributions from all the reconstructed filaments.
Altogether, two preferential bar widths $\simeq 0.7\,{\rm pc}$ and $\simeq 0.9\,{\rm pc}$ emerge from our analysis, though it is not clear whether these are specific to G82 and G202 or whether they could have more generality.
Uncovering general trends in the preferential bar widths would require a complete statistical analysis over many more {\it Herschel} fields.

For each {\it Herschel} field, we can integrate the curve $(N_{\rm pix}/N_{\rm map})$ versus $W_{\rm b}^\star$ (solid line) over $W_{\rm b}^\star$ to obtain the fraction of the map covered by reconstructed filaments with bar width in the considered range ($\simeq [0.04, 0.2]\,{\rm pc}$ for G210 and G300 and $\simeq [0.2, 1.1]\,{\rm pc}$ for G82 and G202).
The result is $\simeq 33\%$ for G210, $\simeq 25\%$ for G300, $\simeq 51\%$ for G82, and $\simeq 55\%$ for G202.
Hence, between one quarter and one half of the {\it Herschel} maps are found to be covered by reconstructed filaments, in agreement with the visual impression left by the reconstructed maps in Appendix~\ref{sec:summary_results_coldensity}.
Let us emphasize, though, that these fractions are probably highly dependent on the angular resolution of the map, in the sense that a given field mapped at higher resolution might be expected to appear more extensively covered by filaments.

It is also interesting to examine how the ${\rm H_2}$ column density varies with bar width in the four {\it Herschel} fields (see the four panels of Fig.~\ref{fig:NH2_vs_Wb}).
In each {\it Herschel} field, we first consider all the pixels whose most significant bar width is $W_{\rm b}^\star$ and calculate their average column density, $\overline{N_{\rm H_2}}$, as a function of $W_{\rm b}^{\star}$ (solid curve).
We then consider all the pixels belonging to a reconstructed filament of bar width $W_{\rm b}$ and calculate their average column density, $\overline{N_{\rm H_2}}$, as a function of $W_{\rm b}$ (dashed curve).
When all the reconstructed filaments are included (dashed curve), the average column density clearly increases with increasing bar width, which means that larger filaments tend to have higher column densities.
When only the most significant filaments are included (solid curve), this trend becomes much weaker (in G82 and G202) or even disappears entirely (in G210 and G300).
What emerges instead are pronounced peaks at intermediate values of $W_{\rm b}^\star$.
The associated high values of $\overline{N_{\rm H_2}}$ suggest that the corresponding intermediate filaments are actually part of larger filaments.
Thus, this figure could indicate the existence of special bar widths at which crests and strands inside large filaments with high column densities are more significant than their enclosing filaments.

\subsection{Filament relative orientations: variations with bar width}
\label{sec:filament_relative_orientation_scale_dependence}

\subsubsection{G82}
\label{sec:filament_relative_orientation_scale_dependence_G82}

\begin{figure*}
    \centering
    \includegraphics[width=\textwidth]{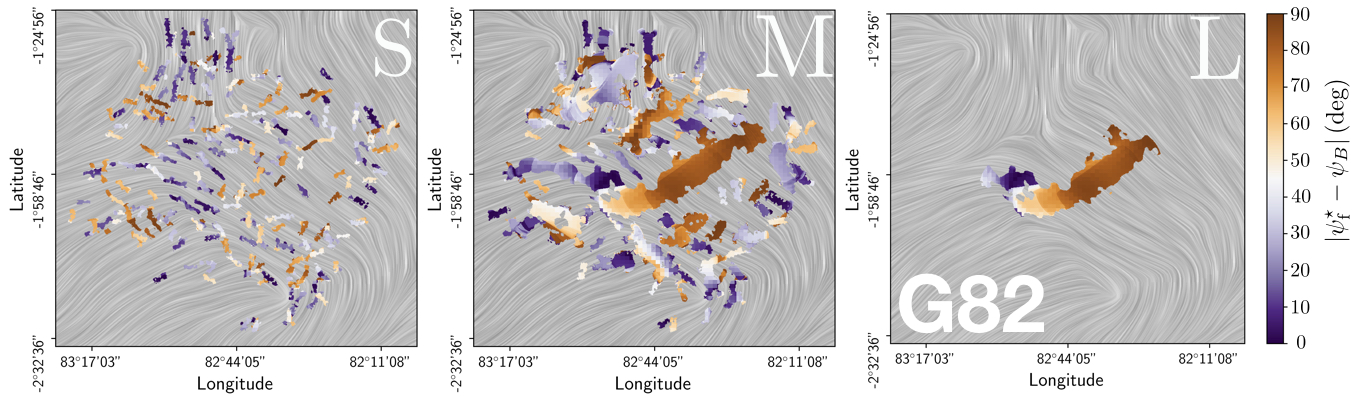}
    \caption{Absolute value of the relative orientation angle between the reconstructed filaments and the magnetic field, $\vert\psi_{\rm f}^{\star} - \psi_B\vert$, in the {\it Herschel} G82 field, for the three ranges of bar width defined in Sect.~\ref{sec:filament_relative_orientation_scale_dependence_G82}: Small ({\bf left}), Medium ({\bf middle}), and Large ({\bf right}). The magnetic field orientation is visualized with LIC in greyscale in the background.}
    \label{fig:relative_orientation_map_G82}
\end{figure*}

\begin{figure}
    \centering
    \includegraphics[width=\columnwidth]{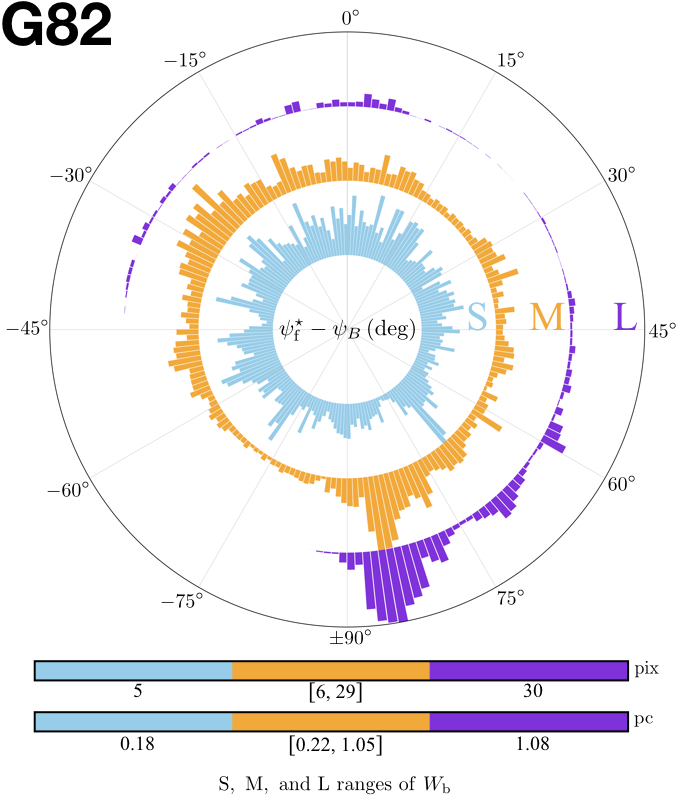}
    \caption{Normalized histograms of relative orientation (HROs) between the reconstructed filaments and the magnetic field, $(N_{\rm pix}/(N_{\rm pix})_{\rm max})$
    versus $(\psi_{\rm f}^{\star} - \psi_B)$, in polar representation over $[-90^\circ,+90^\circ]$, in the {\it Herschel} G82 field, for the three ranges of bar width defined in Sect.~\ref{sec:filament_relative_orientation_scale_dependence_G82}: Small (blue), Medium (orange), and Large (purple).}
    \label{fig:relative_orientation_1D_G82}
\end{figure}

We define three ranges of bar width: the Small (S) range includes a single bin at the smallest bar width, $W_{\rm b}=0.18\,{\rm pc}$, the Large (L) range includes a single bin at the largest bar width, $W_{\rm b}=1.08\,{\rm pc}$, and the Medium (M) range includes all the intermediate bar widths, $W_{\rm b}=[0.22, 1.04]\,{\rm pc}$. In Sect.~\ref{sec:FilDReaMS_overview}, we explained how to reconstruct physical filaments with a given bar width, $W_{\rm b}$, and we defined at each pixel a filament orientation angle, $\psi_{\rm f}^{\star}$, for that bar width.
We now apply this procedure to the S, M, and L ranges,
thereby obtaining S, M, and L filaments, respectively.
S and L filaments are simply the reconstructed filaments with the smallest and largest bar widths, respectively.
M filaments are formed by the superposition of all the reconstructed filaments of intermediate bar widths, and the associated filament orientation angle at each pixel is the filament orientation angle derived for the most significant intermediate bar width.

The left, middle, and right panels of Fig.~\ref{fig:relative_orientation_map_G82} show the absolute value of the relative orientation angle, $\vert\psi_{\rm f}^{\star} - \psi_B\vert$, inside the S, M, and L filaments, respectively, with the magnetic field orientation plotted with Linear Integral Convolution (LIC) in greyscale in the background. The derived values of $\vert\psi_{\rm f}^{\star} - \psi_B\vert$ are in good agreement with the relative orientations inferred from a direct by-eye inspection. The S and M maps exhibit a broad range of relative orientations. In contrast, the L map is dominated by one big filament, which is oriented at $\approx 90^{\circ}$ from the magnetic field -- except for a short portion beyond the kink in the lower part, which is nearly parallel to the magnetic field.

These general trends appear more clearly and in more detail in the three normalized HROs, $(N_{\rm pix}/(N_{\rm pix})_{\rm max})$ versus $(\psi_{\rm f}^{\star} - \psi_B)$,
plotted in polar representation over $[-90^\circ,+90^\circ]$ in Fig.~\ref{fig:relative_orientation_1D_G82}.
The HRO of S filaments (blue) covers the entire range of relative orientation, with a slight asymmetry that favors small angles. 
At the other extreme, the HRO of L filaments (purple) is strongly dominated by a pronounced peak in relative orientation near $90^{\circ}$. Between these two extremes, the HRO of M filaments (orange) exhibits both trends: it covers almost the entire range of relative orientation and has a pronounced peak near $90^{\circ}$; it also has a weaker and broader peak near $-25^{\circ}$.

\subsubsection{G210}
\label{sec:filament_relative_orientation_scale_dependence_G210}

\begin{figure*}
    \centering
    \includegraphics[width=\textwidth]{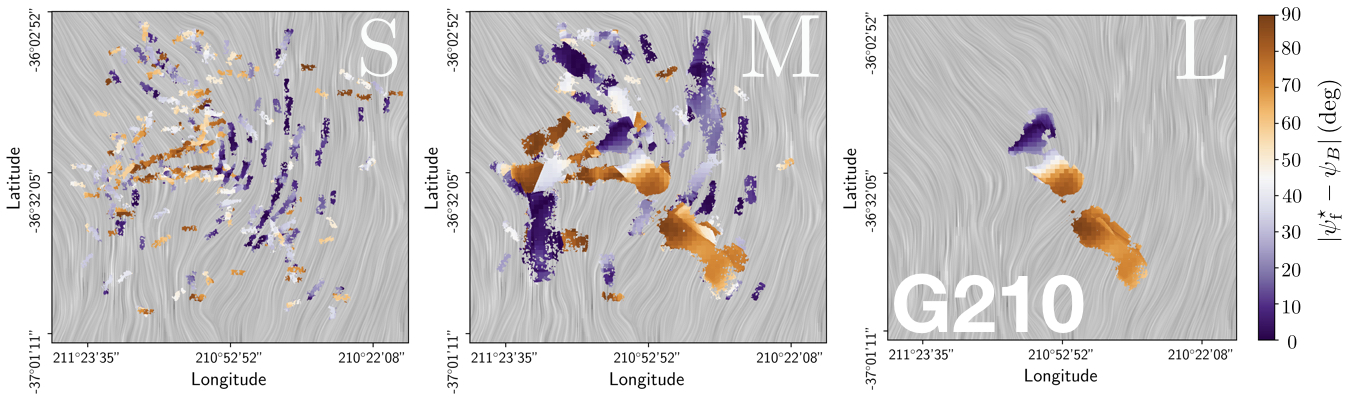}
    \caption{Same as Fig.\ref{fig:relative_orientation_map_G82}, but for the {\it Herschel} G210 field.}
    \label{fig:relative_orientation_map_G210}
\end{figure*}

\begin{figure}
    \centering
    \includegraphics[width=\columnwidth]{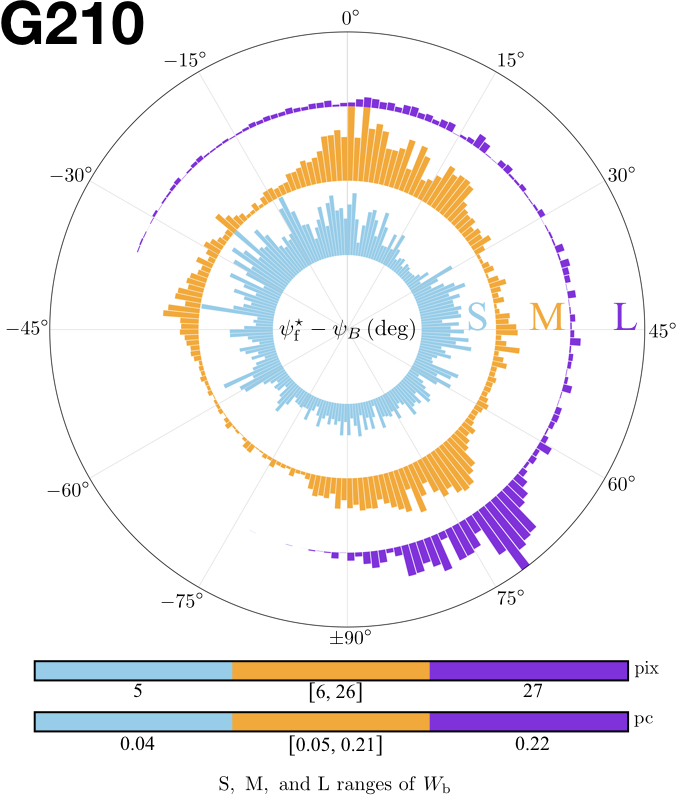}
    \caption{Same as Fig.\ref{fig:relative_orientation_1D_G82}, but for the {\it Herschel} G210 field.}
    \label{fig:relative_orientation_1D_G210}
\end{figure}

We define again S, M, and L filaments, but now with $W_{\rm b}=0.04\,{\rm pc}$, $W_{\rm b}=[0.05, 0.21]\,{\rm pc}$, and $W_{\rm b}=0.22\,{\rm pc}$, respectively.
The S, M, and L filaments, together with their $\vert\psi_{\rm f}^{\star} - \psi_B\vert$, are plotted in the left, middle, and right panels of Fig.~\ref{fig:relative_orientation_map_G210}.
S filaments have again a broad range of relative orientations, with a slight asymmetry in favor of small angles, but they form a more structured spatial pattern than in G82 (Fig.~\ref{fig:relative_orientation_map_G82}): a whole group of filaments on the right is nearly parallel to the magnetic field, while a group on the left is nearly perpendicular.
L filaments reduce to a pair of filaments threaded along a same straight line, which makes a large angle ($\simeq 70^{\circ} - 85^{\circ}$) to the magnetic field
-- except in the uppermost portion of the upper filament, where field lines turn to become nearly parallel to the filament.
M filaments for the most part are either nearly parallel ($\vert\psi_{\rm f}^{\star} - \psi_B\vert \simeq 0^{\circ} - 20^{\circ}$) or nearly perpendicular ($\vert\psi_{\rm f}^{\star} - \psi_B\vert \simeq 70^{\circ} - 90^{\circ}$) to the magnetic field, and they too show more spatial coherence than in G82.

These general trends are nicely confirmed by the HROs plotted in Fig.~\ref{fig:relative_orientation_1D_G210}, which in addition provide more quantitative information on the relevant angular ranges.

\subsubsection{The four {\it Herschel} fields}
\label{sec:filament_relative_orientation_scale_dependence_all_fields}

The S, M, and L maps as well as the HROs of the four {\it Herschel} fields are plotted in their respective sheets in Appendix~\ref{sec:summary_results_coldensity}.

G300 is qualitatively similar to G210. 
Its S, M, and L ranges are $W_{\rm b}=0.04\,{\rm pc}$, $W_{\rm b}=[0.05, 0.16]\,{\rm pc}$, and $W_{\rm b}=0.17\,{\rm pc}$, respectively.
The main difference is that the S and M maps of G300 follow a common global pattern. In the M map, the biggest filaments in the middle are nearly perpendicular to the magnetic field, while the smaller filaments on either side are more nearly parallel.
This global pattern is reflected in the S map, which contains two overlapping families of filaments with nearly orthogonal orientations.
A more minor difference is that the HRO of S filaments in G300 has a stronger asymmetry toward small angles.

G202 also has many similarities with the other three {\it Herschel} fields.
Its S, M, and L ranges are $W_{\rm b}=0.22\,{\rm pc}$, $W_{\rm b}=[0.27, 1.07]\,{\rm pc}$, and $W_{\rm b}=1.11\,{\rm pc}$, respectively.
What clearly sets it apart is that its L filaments are at small angles ($\vert\psi_{\rm f}^{\star} - \psi_B\vert \lesssim 30^{\circ}$) to the magnetic field.
M filaments automatically follow a similar trend.
S filaments, which already show a slight preference for small angles in the other {\it Herschel} fields, only have this preference slightly enhanced in G202.
Altogether, the three types of filaments tend to be more parallel than perpendicular to the magnetic field, and this trend gradually increases from S to M to L filaments.

When the four {\it Herschel} fields are considered together, a few general trends emerge in each of the S, M, and L ranges,
even though these ranges do not refer to the same linear scales for the different fields (see Fig.~\ref{fig:NH2_vs_Wb}):
\begin{itemize}
    \item S filaments are numerous and observed at all relative orientations, with a slight general trend toward alignment parallel to the magnetic field. This trend becomes increasingly noticeable along the sequence G82, G210, G300, G202. \\
    \item The L map is dominated by one or two filaments covering a restricted range of relative orientations, which is close to $90^{\circ}$ in G82, G210, and G300, and broadly around $0^{\circ}$ in G202. \\
    \item M filaments are observed at most relative orientations, with a strong inclination toward the range covered by L filaments.
\end{itemize}
An important conclusion of this first analysis is that filaments of different widths align differently with respect to the magnetic field.

\subsection{Filament relative orientations: variations with column density}
\label{sec:filament_relative_orientation_variation_NH2}

\subsubsection{G82}
\label{sec:filament_relative_orientation_variation_NH2_G82}

\begin{figure*}
    \centering
    \includegraphics[width=\textwidth]{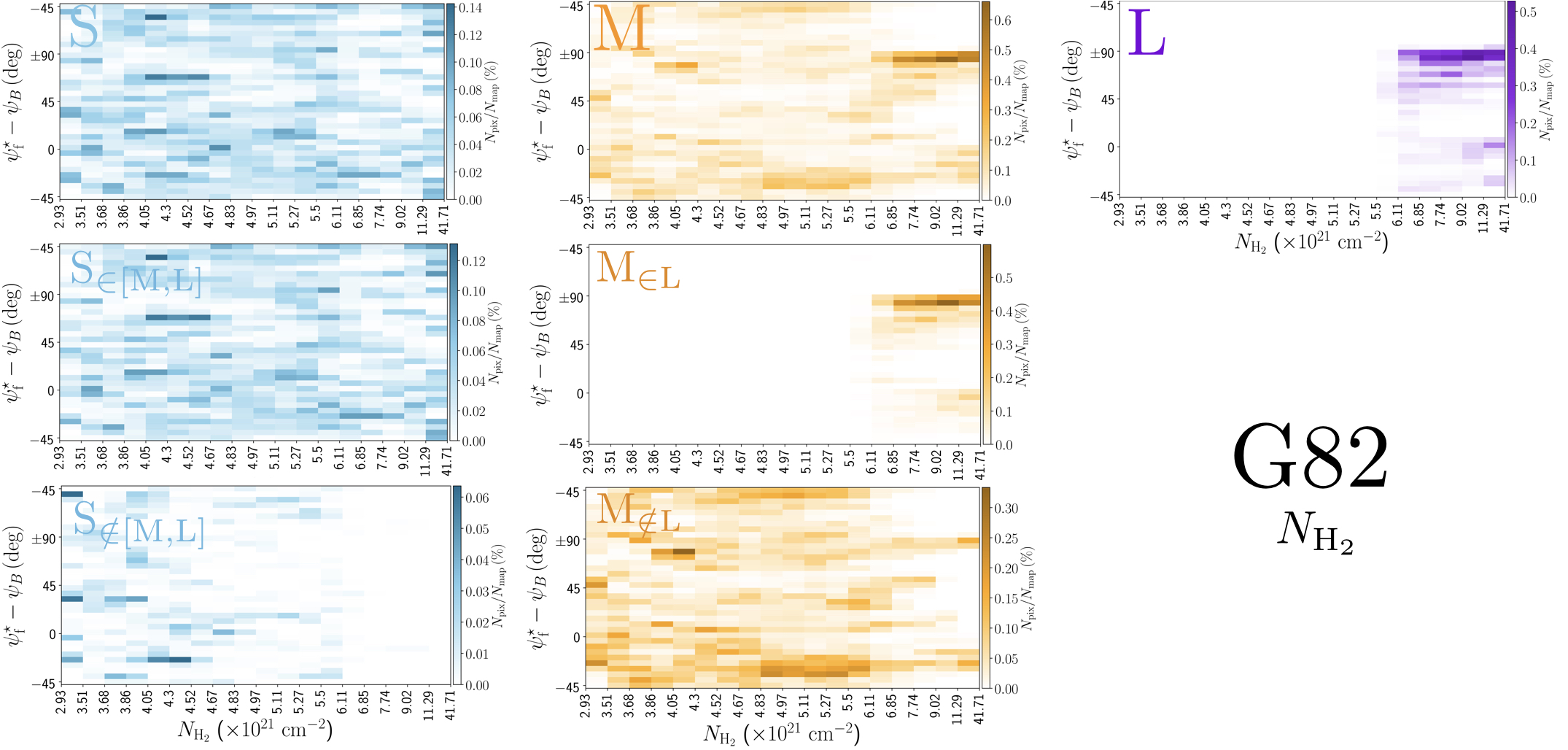}
    \caption{2D histograms of the relative orientation angle between the reconstructed filaments and the magnetic field, $(\psi_{\rm f}^{\star} - \psi_B)$, as a function of ${\rm H_2}$ column density, $N_{\rm H_2}$, in the {\it Herschel} G82 field, for the three ranges of bar width defined in Sect.~\ref{sec:filament_relative_orientation_scale_dependence_G82}: Small ({\bf left}, blue), Medium ({\bf middle}, orange), and Large ({\bf right}, purple).
    {\bf Top}: All of the S, M, and L filaments.
    {\bf Middle}: S and M filaments that belong to larger filaments.
    {\bf Bottom}: S and M filaments that do not belong to larger filaments.}
    \label{fig:relative_orientation_2D_G82}
\end{figure*}

We now inquire into a possible correlation between the relative orientation angle, $(\psi_{\rm f}^{\star} - \psi_B)$, and the {\it Herschel} column density, $N_{\rm H_2}$, for the S, M, and L filaments separately.
Considering only pixels that belong to at least one reconstructed filament, we divide the full range of $N_{\rm H_2}$ into 18 bins containing the same number of pixels. For each of the S, M, and L sets of filaments, we construct the HRO in every $N_{\rm H_2}$ bin and we combine the 18 HROs into the 2D histograms $(N_{\rm pix}/N_{\rm map})$ versus $(N_{\rm H_2}, \psi_{\rm f}^{\star} - \psi_B)$ plotted in the top row of Fig.~\ref{fig:relative_orientation_2D_G82}. Also plotted in Fig.~\ref{fig:relative_orientation_2D_G82} are the 2D histograms of the S filaments that belong (S$_{\rm \in [M, L]}$, middle left panel) or do not belong (S$_{\rm \notin [M, L]}$, bottom left panel) to larger (M or L) filaments as well as the 2D histograms of the M filaments that belong (M$_{\rm \in L}$, central panel) or do not belong (M$_{\rm \notin L}$, bottom middle panel) to L filaments.
The computation of which (S or M) filaments belong to larger filaments is performed on a pixel-by-pixel basis.

In the left column of Fig.~\ref{fig:relative_orientation_2D_G82}, the top panel confirms that S filaments exist at all relative orientations, and shows that this is true at all column densities; the slight asymmetry in favor of small angles detected in the blue HRO of Fig.~\ref{fig:relative_orientation_1D_G82} is hardly noticeable.
The middle and bottom panels, for their part, tell us that (1) the above conclusions directly apply to S$_{\rm \in [M, L]}$ filaments, which represent the vast majority of S filaments, and (2) the few S$_{\rm \notin [M, L]}$ filaments are predominantly found at low column densities.

The rightmost panel of Fig.~\ref{fig:relative_orientation_2D_G82} confirms that L filaments have a strong preference for relative orientations $\simeq 90^{\circ}$. It now clearly appears that this preference applies only at high column densities, where all the L filaments are found.
The latter actually reduce to one big filament, as already mentioned in connection with the right panel of Fig.~\ref{fig:relative_orientation_map_G82}. 

In the middle column of Fig.~\ref{fig:relative_orientation_2D_G82}, the top panel confirms that M filaments exist at nearly all relative orientations, with a marked preference for $\simeq 90^{\circ}$.
We now see that this preference applies mostly at high column densities.
Furthermore, the middle and bottom panels bring to light a clear dichotomy between M$_{\rm \in L}$ filaments, which are almost exclusively found at high column densities, with relative orientations $\simeq 90^{\circ}$, and M$_{\rm \notin L}$ filaments, which are mostly found at low-to-intermediate column densities, with a slight clustering around $-25^{\circ}$.
The M$_{\rm \in L}$ histogram is strikingly similar to the L histogram, which can be explained by the fact that the L filament in the right panel of Fig.~\ref{fig:relative_orientation_map_G82} has an almost identical M counterpart in the middle panel.

\subsubsection{G210}
\label{sec:filament_relative_orientation_variation_NH2_G210}

\begin{figure*}
    \centering
    \includegraphics[width=\textwidth]{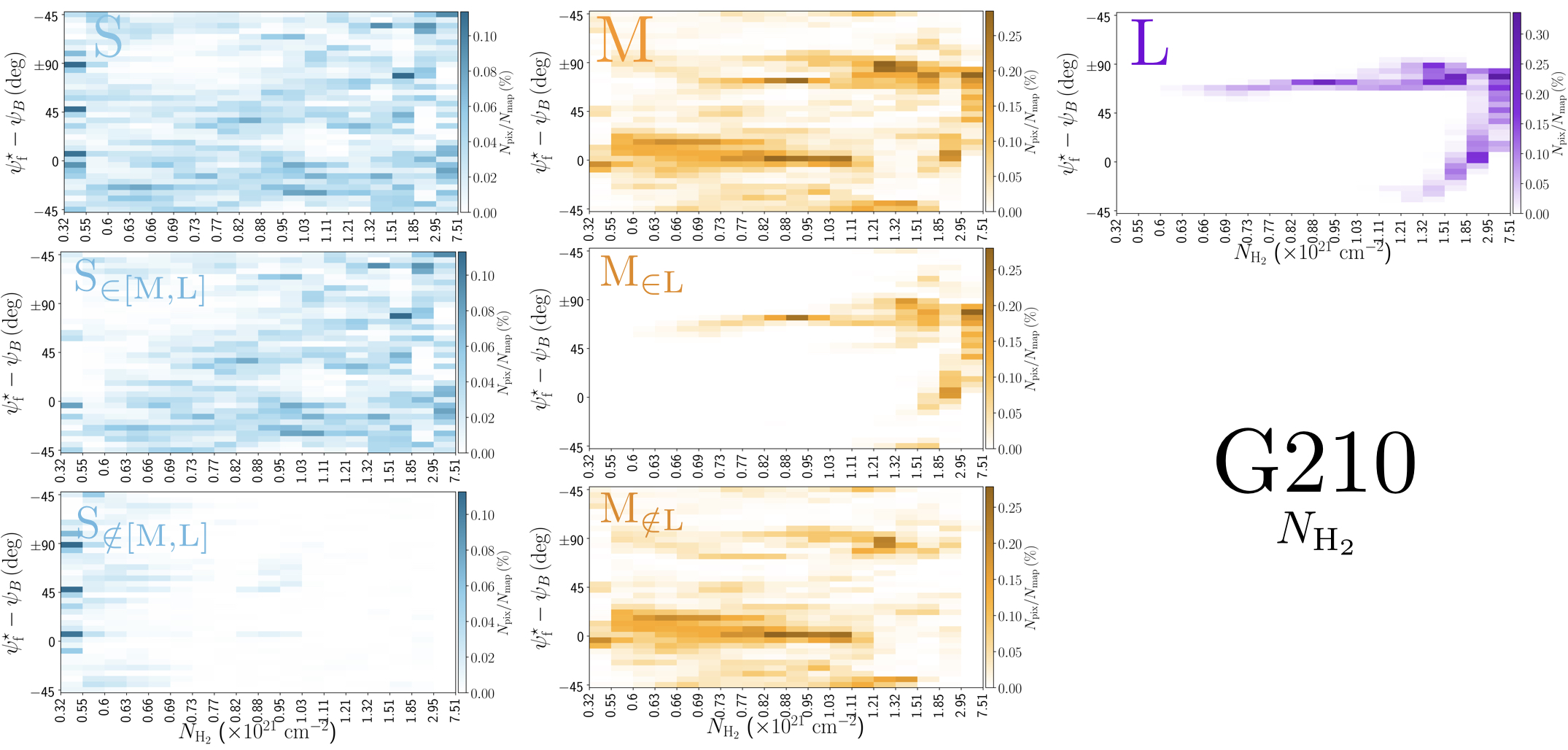}
    \caption{Same as Fig.~\ref{fig:relative_orientation_2D_G82}, but for the {\it Herschel} G210 field.}
    \label{fig:relative_orientation_2D_G210}
\end{figure*}

Following the same procedure as for G82 (Sect.~\ref{sec:filament_relative_orientation_variation_NH2_G82}), we obtain the 2D histograms $(N_{\rm pix}/N_{\rm map})$ versus $(N_{\rm H_2}, \psi_{\rm f}^{\star} - \psi_B)$ displayed in Fig.~\ref{fig:relative_orientation_2D_G210}.

The S filaments (left column) show nearly the same trends as in G82.
Two minor differences are that (1) the slight asymmetry in favor of small relative orientation angles detected in the blue HRO of Fig.~\ref{fig:relative_orientation_1D_G210} is now visible at low column densities, especially in the S$_{\rm \in [M, L]}$ histogram, and (2) the few S$_{\rm \notin [M, L]}$ filaments are almost exclusively found at low column densities and at specific relative orientations ($\simeq 10^{\circ}$, $50^{\circ}$, and $90^{\circ}$).

For L filaments (right column), the broad peak at $\simeq 70^{\circ} - 85^{\circ}$ observed in the purple HRO of Fig.~\ref{fig:relative_orientation_1D_G210} is now seen to arise from a wide range of intermediate-to-high column densities.
In contrast, the few pixels with relative orientations between $\simeq -35^{\circ}$ and $70^{\circ}$ are all concentrated at the highest column densities.
Remember that these pixels belong to the uppermost portion of the upper filament in the right panel of Fig.~\ref{fig:relative_orientation_map_G210}, where field lines undergo significant bending.

For M filaments (middle column), the two preferential relative orientations $\simeq 0^{\circ} - 20^{\circ}$ and $\simeq 70^{\circ} - 90^{\circ}$ emerging from the orange HRO of Fig.~\ref{fig:relative_orientation_1D_G210} are now seen to apply mostly to low-to-intermediate and intermediate-to-high column densities, respectively.
Roughly speaking, the former pertain to the M$_{\rm \notin L}$ filaments and the latter to the M$_{\rm \in L}$ filaments.
The strong similarity between the M$_{\rm \in L}$ and L histograms can again be explained by the morphological resemblance between L filaments and their enclosed M filaments (see right and middle panels of Fig.~\ref{fig:relative_orientation_map_G210}).

\begin{figure}
    \centering
    \includegraphics[width=\columnwidth]{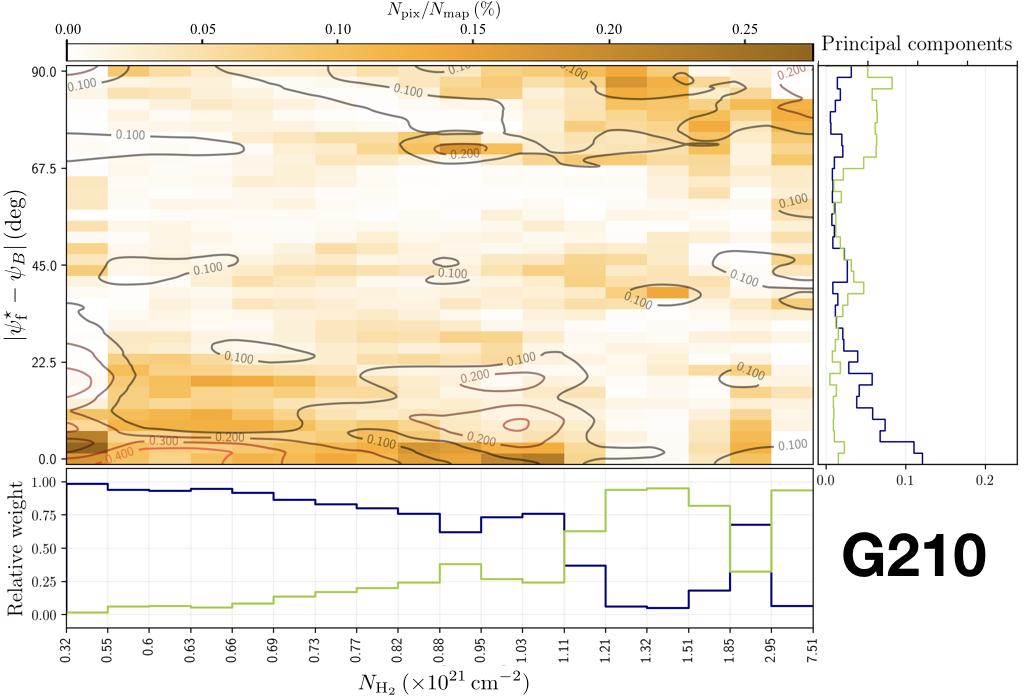}
    \caption{Application of the NMF and PCA methods (introduced in Sect.~\ref{sec:application_orientation}) to the M filaments of the {\it Herschel} G210 field.
    {\bf Top left:} 2D histogram of the relative orientation angle (in absolute value) between M filaments and the magnetic field, $\vert \psi_{\rm f}^{\star} - \psi_B \vert$, as a function of ${\rm H_2}$ column density, $N_{\rm H_2}$.
    {\bf Top right:} Profiles of the two principal components derived with NMF as functions of $\vert \psi_{\rm f}^{\star} - \psi_B \vert$.
    {\bf Bottom:} Relative weights of the two principal components as functions of $N_{\rm H_2}$. Contour lines of the residual of the reconstruction with the two principal components are overplotted on the 2D histogram.}
    \label{fig:PCA_NMF_G210}
\end{figure}

We now focus on the 2D histogram of M filaments and use the NMF and PCA methods introduced in Sect.~\ref{sec:application_orientation} to determine the column density at which the transition in $\vert\psi_{\rm f}^{\star} - \psi_B\vert$, from $\simeq 0^{\circ} - 20^{\circ}$ to $\simeq 70^{\circ} - 90^{\circ}$, occurs.
The result is displayed in Fig.~\ref{fig:PCA_NMF_G210}.
The 2D histogram $(N_{\rm pix}/N_{\rm map})$ versus $(N_{\rm H_2},| \psi_{\rm f}^{\star} - \psi_B|)$ (directly obtained from the top middle panel of Fig.~\ref{fig:relative_orientation_2D_G210}) is plotted in the top panel.
The residual of the reconstruction with the first two principal components derived with NMF (2D histogram minus the sum of the two principal components) is overplotted with contour lines.
The first two principal components together have an estimated reconstruction capability of 65$\,\%$. Their profiles as functions of $\vert \psi_{\rm f}^{\star} - \psi_B \vert$ are shown in the right panel of Fig.~\ref{fig:PCA_NMF_G210}, where the blue and green components mostly represent filaments that are approximately parallel and perpendicular, respectively, to the magnetic field.
The relative weights of the two components as functions of column density are shown in the bottom panel.
The "parallel component" dominates at $N_{\rm H_2} < 1.1\times10^{21}\,$cm$^{-2}$, while the "perpendicular component" dominates at $N_{\rm H_2} > 1.1\times10^{21}\,$cm$^{-2}$ -- except in the bin $\left[1.85, 2.95\right]\times10^{21}\,$cm$^{-2}$, which is dominated by the "parallel component".
This nearly parallel orientation at high column density comes from
the uppermost portion of the M filament associated with the upper L filament in the right panel of Fig.~\ref{fig:relative_orientation_map_G210}.
What can be retained from the application of the NMF and PCA methods is that the relative orientation of filaments undergoes a transition from mostly parallel to mostly perpendicular to the magnetic field at a column density $N_{\rm H_2} \simeq 1.1\times10^{21}\,$cm$^{-2}$.

\subsubsection{The four {\it Herschel} fields}
\label{sec:filament_relative_orientation_variation_NH2_all_fields}

The 2D histograms $(N_{\rm pix}/N_{\rm map})$ versus $(N_{\rm H_2}, \psi_{\rm f}^{\star} - \psi_B)$ of the S, M, and L filaments in the four {\it Herschel} fields are displayed in their respective sheets in Appendix~\ref{sec:summary_results_coldensity}.
For compactness, we do not show the 2D histograms of the S$_{\rm \in [M, L]}$ and S$_{\rm \notin [M, L]}$ filaments or those of the M$_{\rm \in L}$ and M$_{\rm \notin L}$ filaments separately, as we did for G82 in Fig.~\ref{fig:relative_orientation_2D_G82} and G210 in Fig.~\ref{fig:relative_orientation_2D_G210}.
However, we did construct and examine the separate 2D histograms for the four {\it Herschel} fields, and we found that they all share the same general properties.

\begin{figure}
    \centering
    \includegraphics[width=\columnwidth]{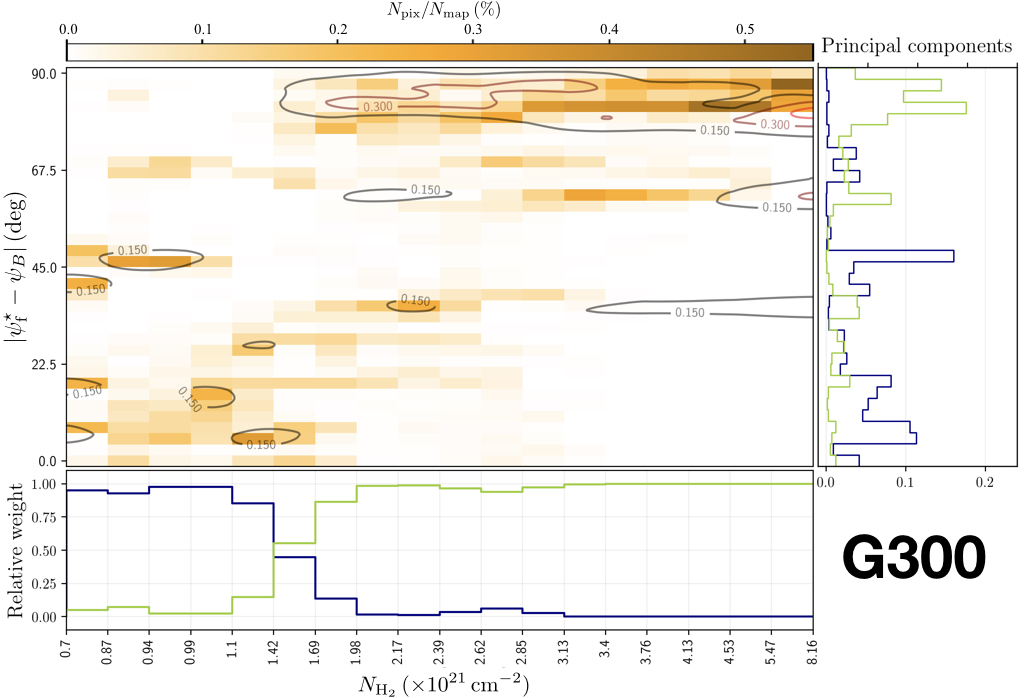}
    \caption{Same as Fig.~\ref{fig:PCA_NMF_G210}, but for the {\it Herschel} G300 field.}
    \label{fig:PCA_NMF_G300}
\end{figure}

G300 has many similarities with G210 (perhaps in part because both are nearby fields), as well as its own peculiarities.
S filaments with low column densities show a clear asymmetry toward small relative orientation angles, and so do S$_{\rm \in [M, L]}$ filaments.
L filaments reduce to a single intermediate-to-high $N_{\rm H_2}$ filament, with relative orientation $\simeq 90^{\circ}$.
M filaments can be divided into two classes: low-to-intermediate $N_{\rm H_2}$ filaments, with relative orientations between $-45^{\circ}$ and $+45^{\circ}$, and intermediate-to-high $N_{\rm H_2}$ filaments, with relative orientations $\approx 90^{\circ}$.
These two classes can be roughly identified with the M$_{\rm \notin L}$ and M$_{\rm \in L}$ filaments, respectively.
They also correspond to the two principal components derived with the NMF method applied to the M filaments, with an estimated reconstruction capability of 71$\,\%$ (see Fig.~\ref{fig:PCA_NMF_G300}).
The bottom panel of Fig.~\ref{fig:PCA_NMF_G300} indicates that the transition between the two components occurs at a column density $N_{\rm H_2} \simeq 1.4\times10^{21}\,$cm$^{-2}$, which turns out to be very close to the transition column density $\simeq 1.1\times10^{21}\,$cm$^{-2}$ obtained for G210.

G202 shows similar trends to the other three {\it Herschel} fields, but only for S, S$_{\rm \in [M, L]}$, S$_{\rm \notin [M, L]}$, and M$_{\rm \notin L}$ filaments.
L and M$_{\rm \in L}$ filaments, which are still found at high column densities, now cover a broad range of relative orientations around $0^{\circ}$.

To sum up, the general trends emerging from the 2D histograms of S, M, and L filaments in the four {\it Herschel} fields are the following:

\begin{itemize}
    \item S filaments are found at all column densities and all relative orientations. Those with low column densities have a general tendency (hardly noticeable in G82, weak in G210 and G202, stronger in G300) to be more parallel than perpendicular to the magnetic field. The small fraction of S filaments present outside larger (M or L) filaments mostly have low column densities, with no preference for parallel alignment. \\
    \item L filaments have high or intermediate-to-high column densities. In G82, G210, and G300, they tend to be nearly perpendicular to the magnetic field, while in G202, they tend to be more nearly parallel. \\
    \item M filaments span the entire ranges of column densities and relative orientations, with however a much more structured distribution than S filaments. M$_{\rm \in L}$ filaments behave very similarly to L filaments.
\end{itemize}

\begin{figure*}
    \centering
    \includegraphics[width=\textwidth]{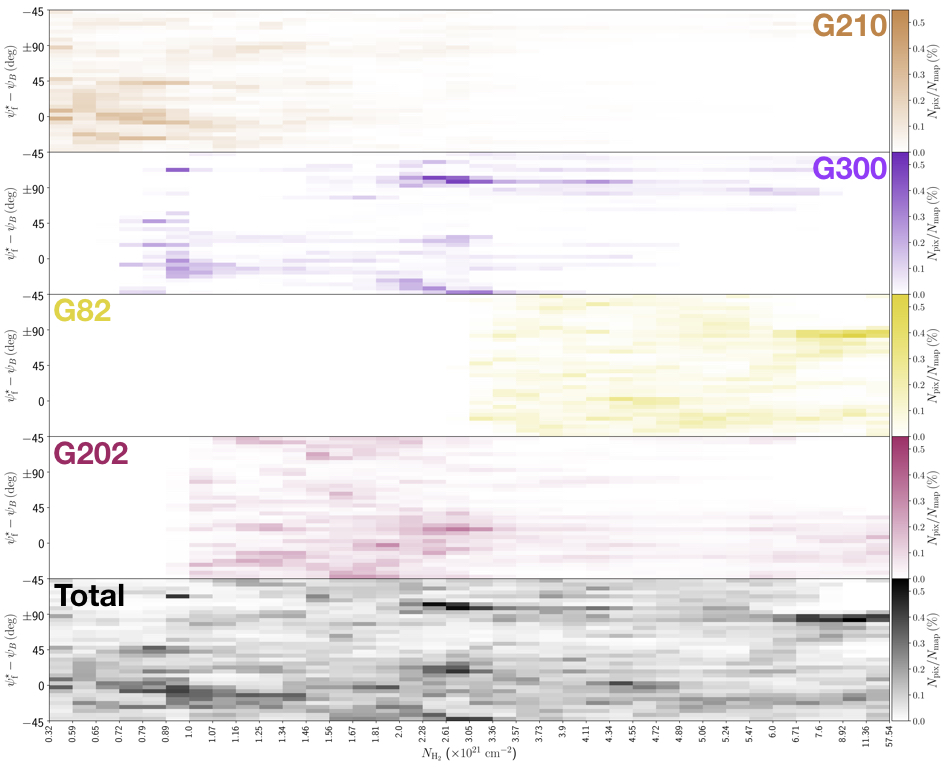}
    \caption{2D histograms of the relative orientation angle between the reconstructed (S, M, and L) filaments and the magnetic field, $(\psi_{\rm f}^{\star} - \psi_B)$, as a function of ${\rm H_2}$ column density, $N_{\rm H_2}$.
    {\bf Top four panels:} Individual histograms of the four {\it Herschel} fields separately. 
    {\bf Bottom panel:} Combined histogram of the four {\it Herschel} fields together.}
    \label{fig:preferential_RO_fields}
\end{figure*}

If we now consider S, M, and L filaments together, we can derive the complete 2D histograms $(N_{\rm pix}/N_{\rm map})$ versus $(N_{\rm H_2}, \psi_{\rm f}^{\star} - \psi_B)$ of the four individual {\it Herschel} fields (top four panels of Fig.~\ref{fig:preferential_RO_fields}),
as well as the combined 2D histogram of the four {\it Herschel} fields together (bottom panel).
The full range of $N_{\rm H_2}$, which encompasses the different $N_{\rm H_2}$ ranges of the individual {\it Herschel} fields, is divided into 36 bins with the same sum (over the four fields) of $N_{\rm pix}/N_{\rm map}$.
Clearly, for each {\it Herschel} field, the complete histogram (Fig.~\ref{fig:preferential_RO_fields}) does not closely resemble any of the S, M, or L histogram (Appendix~\ref{sec:summary_results_coldensity}).
This results from a combination of two different factors:
(1) Each pixel retained in the complete histogram is assigned the orientation angle of the most significant filament passing through it, which can now be a S, M, or L filament.
(2) The $N_{\rm H_2}$ grid based on the four {\it Herschel} fields is very different from the individual $N_{\rm H_2}$ grids.

The combined histogram (bottom panel of Fig.~\ref{fig:preferential_RO_fields}), which is simply the superposition of the four individual complete histograms, tentatively shows a bimodal distribution in relative orientation, with a tendency for low-/high-$N_{\rm H_2}$ filaments to be more nearly parallel/perpendicular to the magnetic field.
The transition from parallel to perpendicular orientation occurs over a range of column densities $\approx [2,3] \times10^{21}\,$cm$^{-2}$.
However, this range is biased upward by G202, which contains mostly parallel filaments throughout its $N_{\rm H_2}$ range.
If G202 were excluded from the combined histogram, the transition would occur at $N_{\rm H_2} \approx (1-2) \times10^{21}\,$cm$^{-2}$, consistent with the values obtained earlier for G210 ($\simeq 1.1\times10^{21}\,$cm$^{-2}$) and G300 ($\simeq 1.4\times10^{21}\,$cm$^{-2}$).
Here, too, a more accurate estimate of the transition column density would require a complete statistical analysis over a large number of {\it Herschel} fields.

\begin{table*}\centering 
\caption{
Summary of the main results obtained for the reconstructed filaments in each of the four {\it Herschel} fields.}
\begin{threeparttable}
\begin{tabular}{m{2.3cm} | m{0.01cm} m{2cm} m{2cm} m{0.01cm} | m{2cm} m{4.5cm} m{1.cm} m{0.01cm}}
\toprule\toprule
\textbf{\textit{Herschel} field}
& & \multicolumn{2}{c}{Most prevalent widths $[{\rm pc}]$} & & \multicolumn{3}{c}{Preferential relative orientations}\\ 

Summary figure
&
& \centering $W_{\rm b}^{\star {\rm peak}}$
& \centering 2$R_{\rm flat}\,(p\!=\!2.2)$
& & \centering S & \centering M & \centering L &\\ 

\midrule

\textbf{\large G210} &     & \centering $0.04\pm0.01$ & \centering $0.07\pm0.03$ & & $[0.04]\,{\rm pc}$ & $[0.05, 0.21]\,{\rm pc}$                                       & $[0.22]\,{\rm pc}$ &\\
Fig.~\ref{fig:G210_plots} & & \centering $0.15\pm0.02$ & \centering $0.23\pm0.05$ & & $\diagup$          & $\parallel$ at $N_{\rm H_2} < 1.1\times10^{21}\,{\rm cm^{-2}}$ & $\perp$ &           \\
                          & & \centering $0.18\pm0.03$ & \centering $0.27\pm0.05$ & &                    & $\perp$ at $N_{\rm H_2} > 1.1\times10^{21}\,{\rm cm^{-2}}$     & &                   \\
                          & & \centering $0.22\pm0.03$ & \centering $0.33\pm0.06$ & &                    & $N_{\rm H_2}$ transition (Fig.~\ref{fig:PCA_NMF_G210})         & &                   \\
                     
& & & & & & \\

\textbf{\large G300}       & & \centering $0.04\pm0.01$ & \centering $0.06\pm0.02$ & & $[0.04]\,{\rm pc}$ & $[0.05, 0.16]\,{\rm pc}$                                       & $[0.17]\,{\rm pc}$ &\\
Fig.~\ref{fig:G300_plots} & & \centering $0.06\pm0.01$ & \centering $0.09\pm0.03$ & & $\diagup$          & $\parallel$ at $N_{\rm H_2} < 1.4\times10^{21}\,{\rm cm^{-2}}$ & $\perp$ &           \\
                            & & \centering $0.17\pm0.03$ & \centering $0.25\pm0.06$ & &                    & $\perp$ at $N_{\rm H_2} > 1.4\times10^{21}\,{\rm cm^{-2}}$     & &                   \\
                            & &        &        &        &                    & $N_{\rm H_2}$ transition (Fig.~\ref{fig:PCA_NMF_G300})       & &                   \\

& & & & & & \\

\textbf{\large G82}      & & \centering $0.18\pm0.01$ & \centering $0.29\pm0.06$ & & $[0.18]\,{\rm pc}$ & $[0.22, 1.05]\,{\rm pc}$                                   & $[1.08]\,{\rm pc}$ &\\
Fig.~\ref{fig:G82_plots} & & \centering $0.32\pm0.02$ & \centering $0.49\pm0.07$ & & $\diagup$          & $\perp$ at $N_{\rm H_2} > 6.1\times10^{21}\,{\rm cm^{-2}}$ & $\perp$ &           \\
                         & & \centering $0.72\pm0.04$ & \centering $1.05\pm0.11$ & &                    &                                                            & &                   \\
                         & & \centering $0.90\pm0.05$ & \centering $1.31\pm0.12$ & &                    &                                                            & &                   \\
                         & & \centering $1.08\pm0.06$ & \centering $1.57\pm0.14$ & &                    &                                                            & &                   \\
                     
& & & & & & \\

\textbf{\large G202}      & & \centering $0.22\pm0.03$ & \centering $0.35\pm0.09$ & & $[0.22]\,{\rm pc}$ & $[0.27, 1.07]\,{\rm pc}$                                     & $[1.11]\,{\rm pc}$ &\\
Fig.~\ref{fig:G202_plots} & & \centering $0.71\pm0.09$ & \centering $1.05\pm0.18$ & & $\parallel$      & $\parallel$ at $N_{\rm H_2} > 2.0\times10^{21}\,{\rm cm^{-2}}$ & $\parallel$ &       \\
                          & & \centering $0.97\pm0.13$ & \centering $1.42\pm0.23$ & &                  &                                                                & &                   \\

\bottomrule\bottomrule
\end{tabular}

\begin{tablenotes}
  \item {\bf Column 1:} Name of the {\it Herschel} field and summary figure with the graphical results. %%derived plots.
    {\bf Columns 2--3:} Most prevalent bar widths, $W_{\rm b}^{\star{\rm peak}}$, and corresponding Plummer widths, $2R_{\rm flat}$, for a Plummer power-law index $p=2.2$ (see Fig.~8 and Table~4 of Paper~1).
    {\bf Columns 4--6:} Preferential relative orientations with respect to the magnetic field, for the three ranges of bar width defined in Sect.~\ref{sec:filament_relative_orientation_scale_dependence_G82}: Small (S), Medium (M), and Large (L).
    The ranges in parsecs are given in the first row, and the dominant trends are summarized in the next rows,
    with the symbols $\parallel$, $\perp$, and $\diagup$ meaning preferentially parallel, preferentially perpendicular, and without clear preference, respectively.
\end{tablenotes}
\end{threeparttable}
\label{tab:sum_fields}
\end{table*}

\section{Discussion}
\label{sec:discussion}

\subsection{Filament bar widths}

By applying {\tt FilDReaMS} to the ${\rm H_2}$ column density maps of the four {\it Herschel} fields of our sample, we were able to construct histograms of the most significant bar width, $W_{\rm b}^\star$ (solid line in the top four panels of Fig.~\ref{fig:average_significance_fields}), and from the peaks of the histograms we were able to derive the most prevalent bar widths, $W_{\rm b}^{\star{\rm peak}}$.
The histogram of each {\it Herschel} field is found to contain between 4 and 6 peaks, with the outermost peaks (located at the boundaries) being probably partly artificial.
When the four fields are considered together (bottom panel of Fig.~\ref{fig:average_significance_fields}), no significant peak emerges in the range $\simeq [0.04, 0.2]\,{\rm pc}$ covered by the nearby fields, G210 and G300, whereas two pronounced peaks near $0.7\,{\rm pc}$ and $0.9\,{\rm pc}$ stand out in the range $\simeq [0.2, 1.1]\,{\rm pc}$ covered by the distant fields, G82 and G202.
Our current small-number statistics do not allow us to ascribe any generality to these preferential bar widths.

The most prevalent bar widths, $W_{\rm b}^{\star{\rm peak}}$, derived in our study can cautiously be converted to Plummer widths, $2R_{\rm flat}$, with the help of Fig.~8 and Table~4 of Paper~1. We consider the case of default noise level (i.e. the typical noise level of {\it Herschel} maps) and a Plummer power-law index of $p=2.2$ \citep[median value obtained by][for a sample of 599 filaments including G300]{Arzoumanian_2019}.
However, one has to be aware that the resulting $2R_{\rm flat}$ are just rough estimates obtained very indirectly through a method that is not designed to derive the transverse column density profiles of filaments.
Therefore, it is probably not very meaningful to compare our $2R_{\rm flat}$ to the Plummer widths derived in previous studies, for instance, based on true Plummer-type fits to the transverse column density profiles of filaments all along their lengths \citep{Juvela_GCCIII_2012, Kainulainen_2016, Cox_2016, Arzoumanian_2011, Arzoumanian_2019}.

In the case of G210 and G300, the most prevalent bar widths could potentially be found in the range $W_{\rm b}^{\star{\rm peak}} \simeq [0.04, 0.2]\,{\rm pc}$, corresponding to Plummer widths in the range $2R_{\rm flat} \simeq [0.06, 0.3]\,{\rm pc}$.
In the case of G82 and G202, the relevant ranges are $W_{\rm b}^{\star{\rm peak}} \simeq [0.2, 1.1]\,{\rm pc}$, corresponding to $2R_{\rm flat} \simeq [0.3, 1.6]\,{\rm pc}$.
The two most prevalent bar widths in G82 and G202, $W_{\rm b}^{\star{\rm peak}} \simeq 0.7\,{\rm pc}$ and $0.9\,{\rm pc}$, translate into Plummer widths $2R_{\rm flat} \simeq 1.0\,{\rm pc}$ and $1.3\,{\rm pc}$, respectively.
No other preferential Plummer width is uncovered in our sample.

We do not recover the characteristic width $\sim 0.1\,{\rm pc}$ found in previous studies \citep[][]{Arzoumanian_2011, Arzoumanian_2019, Palmeirim_2013, Cox_2016}.
The existence of this characteristic width was called into question by a re-analysis of the \cite{Arzoumanian_2011, Arzoumanian_2019} methodology \citep{Panopoulou_2017, Panopoulou_2022} as well as by the conclusions of an anisotropic wavelet analysis applied to the {\it Herschel} column-density maps of the nearby Aquila Rift and Polaris Flare \citep{Ossenkopf-Okada_2019}.
Here we do not have enough statistics to settle the debate.
The angular resolution (36") of our column density maps precludes detecting filaments widths $\sim 0.1\,{\rm pc}$ in molecular clouds more distant than $\sim 200\,{\rm pc}$ \citep[][]{Kainulainen_2016}, which means that a filament width $\sim 0.1\,{\rm pc}$ is undetectable in our two distant fields (G82 and G202). For the two nearby fields (G210 and G300), where a filament width $\sim 0.1\,{\rm pc}$ falls within the detectability range, its non-detection could suggest that either these two fields are not particularly representative, or the level of noise is too high for the relation between bar width and Plummer width to be really meaningful, or the $\sim 0.1\,{\rm pc}$ is not a universal characteristic filament width.

{\tt FilDReaMS} is not designed to give information on the transverse column density profile of filaments. However, the appearance of small filaments superposed on larger filaments (see right panel of Fig.~\ref{fig:filaments_G82}, bottom panel of Fig.~\ref{fig:filament_best_sizes_G82}, and similar figures in Appendix~\ref{sec:summary_results_coldensity}) gives some rough, qualitative idea of the internal structure of large filaments. In particular, the fact that a fraction of the small (S) filaments form the crests of larger filaments may provide some indirect support for the often-used Plummer-type profile.

The ramifications and striations observed in Fig.~\ref{fig:filaments_G82}, Fig.~\ref{fig:filament_best_sizes_G82}, and Appendix~\ref{sec:summary_results_coldensity} are similar to those observed in dust-continuum and molecular-line maps \citep[see, for instance,][]{Sugitani_2011,Arzoumanian_2013, Peretto_2013, Saajasto_2017, Palmeirim_2013, Andre_2014}.
They suggest a hierarchical process in filament formation, with small filaments potentially feeding larger filaments.
Such a scenario is consistent with the predictions of numerical simulations \citep[see][]{Balsara_2001,GomezVasquezSemadeni_2014}.

\subsection{Filament orientations to the magnetic field}

\subsubsection{Comparison with previous studies} 
The results obtained for the relative orientations between filaments and the local magnetic field in our sample of {\it Herschel} fields confirm some of the general trends observed in previous studies (as introduced in Sect.~\ref{sec:introduction}), but they also display a variety of behaviors. Globally we find that filaments at high column densities are mostly perpendicular to the magnetic field, while a subset of low column-density filaments (mainly striations) are mostly parallel. In two of our fields (G210 and G300), we find a transition between parallel and perpendicular alignments at ${\rm H_2}$ column densities ($\simeq 1.1\times10^{21}\,$cm$^{-2}$ and $1.4\times10^{21}\,$cm$^{-2}$, respectively), consistent with the results from previous studies \citep[][]{Bracco_2016, Soler_2016, Malinen_2016, Alina_2019}. However, these general results are not systematically observed in our sample: low column-density filaments can cover all relative orientations, and in G202, all filaments, including those at high column densities, tend to be more parallel than perpendicular to the magnetic field.

The main novelty of our methodology is that we now have control over the bar widths of filaments and we can investigate the filament relative orientations in different ranges of bar width, down to the angular resolution of the maps. The HROs of our {\it Herschel} fields show different behaviors for S, M, and L filaments, respectively. In each field, a large number of S filaments are detected at all relative orientations, although with a slight trend toward parallel alignment. The largest filaments are mostly perpendicular and are also associated with the highest column densities. Many S and M filaments are parts of larger filaments, with S filaments forming crests, internal sub-structures (such as strands), or ramifications.
Our method makes it possible to separate them from isolated filaments and to study their relative orientations separately.

In Sect.~\ref{sec:analysis_results}, we presented the results obtained when applying our methodology to $N_{\rm H_2}$ maps of our sample. This enabled us to identify truly material filamentary structures. However, the angular resolution of the column density maps is limited to 36" (the resolution of the $500\,{\rm \mu m}$ {\it Herschel} band). In contrast, the {\it Herschel} intensity maps at $250\,{\rm \mu m}$ offer a better angular resolution (18"), which therefore gives access to thinner filaments. For comparison, we performed the same analysis based on the intensity maps of our sample. The results are displayed in Appendix~\ref{sec:summary_results_intensity} (Figs.~\ref{fig:G210_I_plots}--\ref{fig:G202_I_plots}). The gain in angular resolution clearly leads to an increase in the number of detected filaments in the S range. Indeed, improving the angular resolution allows us to detect filaments that are either thinner or at lower column densities. 
These filaments are expected to be parallel to the magnetic field, and this would explain the observed trend in the 2D histograms. Some of these filaments could potentially be striations, or fibers in the sense defined by \citet{Hacar_2013, Hacar_2018}. Such a result was already noticed in the study by \citet{Clark_2014} when using HI data at higher resolution. In the nearby fields G210 and G300, many more striations and strands are detected; the HROs of S filaments are now dominated by nearly parallel orientations, and their 2D histograms show that this trend prevails mainly at low column densities. In G82, there is an increased number of parallel S filaments at all column densities. In G202, the trends observed in both the HROs and the 2D histograms from the $N_{\rm H_2}$ map are just enhanced. 

We now compare our results to those of previous individual studies of the {\it Herschel} fields of our sample. G210 was studied by \citet[][]{Malinen_2016} using {\it Planck} polarization data at 10' resolution and applying the {\tt RHT} method to the {\it Herschel} $250\,{\rm \mu m}$ intensity map. 
In Paper~1, we applied {\tt FilDReaMS} to the same G210 intensity map in order to compare the filaments detected with {\tt RHT} and {\tt FilDReaMS}, and we concluded
that both methods yield similar filamentary networks, although with a few differences.
Interestingly, we found in the G210 column density map a transition between parallel and perpendicular relative orientations at an ${\rm H_2}$ column density $\simeq 1.1\times10^{21}\,$cm$^{-2}$, close to the value $\simeq 0.8\times10^{21}\,$cm$^{-2}$ obtained by \citet[][]{Malinen_2016} from the G210 intensity map.
Here, {\tt RHT} and {\tt FilDReaMS} perform equally to estimate the transition column density. 
The additional asset of {\tt FilDReaMS} is that it enables us to quantify the bar widths of filaments.
In the case of G210, $W_{\rm b}$ varies from $0.04\,{\rm pc}$ to $0.22\,{\rm pc}$.

We find similar characteristics in the {\it Herschel} G300 field, although the global morphology of the cloud as well as its Galactic environment are quite different. The detected filaments cover a similar range of bar width, and we derive a transition between parallel and perpendicular alignment at a very close column density, $N_{\rm H_2} \simeq 1.4\times10^{21}\,$cm$^{-2}$.
A qualitative comparison with the map of filaments obtained by \citet[][]{Cox_2016} (their Fig.~3, derived from a column density map including G300, at 18" resolution) shows a strong similarity with our S filaments, especially those extracted from the $250\,{\rm \mu m}$ intensity map (see Fig.~\ref{fig:G300_I_plots}). Striations and strands are similarly detected, although the methodologies are quite different. 
Our maps of M and L filaments helpfully complete the view at small scales. As \citet[][]{Cox_2016}, we find that the low column-density strands and striations are oriented perpendicular to the main large filament and are parallel to the magnetic field.

Striations are also detected in the {\it Herschel} G82 field, and found to be oriented perpendicular to the main filament. Based on an analysis of molecular line observations in this field, \citet[][]{Saajasto_2017} evidenced that at least one of these striations is kinematically connected to clumps embedded in the main filament, suggesting mass accretion from the striation onto the main filament. Their study indicates that the main filament is highly fragmented, consistent with our own finding that the S filaments detected as sub-structures of the main filament do not follow any ordered configuration as in G300, but instead are oriented at all angles.    

The {\it Herschel} G202 field displays a different behaviour, with an ordered magnetic field structure at large scale over the whole field. Our study shows that filaments at all scales (S, M, L), including the densest and largest filaments, tend to be parallel to the magnetic field. 
This conclusion is in agreement with the results of \citet[][]{Alina_2021}. Combining {\it Planck} observations of dust polarized emission and CO molecular line tracers, they studied the large-scale magnetic field and its interplay with the gas dynamics in this region. Their analysis reveals a shock region with colliding filaments, and they suggest that the magnetic field remains dragged during the evolution of the cloud, resulting in an orientation parallel to the filaments.

\subsubsection{Physical interpretation}

Physically, filaments arise from an interplay between turbulence, magnetic fields, and self-gravity \citep[see, for instance,][]{Andre_2014, Li_2014, Myers_2017}.
These three processes are scale-dependent, and so is their relative importance.
\citet[][]{Chen_2016} showed with three-dimensional (3D) MHD simulations that turbulence dominates at large (cloud) scales, magnetic fields at intermediate (filament) scales, and self-gravity at small (core) scales.
This conclusion also emerges from a simple toy model of a homogeneous cloud of size $\ell$: if interactions between the three processes are ignored, turbulent energy density varies as $\ell$ (Larson's law), magnetic energy density as $\ell^{0}$, and self-gravitational energy density as $\ell^{-1}$ (C-Y. Chen's talk at the Sofia on-line workshop {\it 'Magnetic Fields and the structure of the filamentary Interstellar Medium', June 2021}).
This ordering has direct implications for the formation of filaments and for their orientations with respect to the ambient magnetic field.

At early stages (large scales, low densities), turbulence dominates and leads to the formation of low-density structures through compression \citep{Padoan_2001} and shear \citep{Hennebelle_2013_1} of the gas and the frozen-in field lines; these structures tend to be elongated parallel to the magnetic field.
Field-aligned striations can also form though the non-linear coupling of MHD waves in the presence of density inhomogeneities \citep{Tritsis_2016}.
At late stages (small scales, high densities), self-gravity becomes dominant and causes gas to contract preferentially along field lines; the resulting high-density structures tend to be elongated perpendicular to the magnetic field
\citep[e.g.,][]{Chen_2016, Seifried_2020, Girichidis_2021}.

\citet{Soler_2017} showed that a system in ideal-MHD turbulence naturally tends to evolve towards a configuration where iso-density contours (and hence density structures) are either parallel or perpendicular to the magnetic field.
They also found that, in the presence of a relatively strong magnetic field, compressive motions, resulting from either gravitational collapse or converging flows, can produce a transition from mostly parallel at low column densities to mostly perpendicular at higher column densities.

One can think of several reasons why our observational results do not consistently conform to the theoretical predictions and why different clouds exhibit different behaviors. There are at least two strictly physical reasons.
(1) The transition from preferentially parallel to preferentially perpendicular filaments
occurs at a critical density where self-gravity takes over the dominant role from magnetic fields.
This critical density is strongly dependent on the initial physical conditions (density, ionization fraction, magnetic field, turbulence...) in the parent molecular cloud, which in turn vary significantly from cloud to cloud.
(2) The magnetic field provides a natural reference direction to measure the orientations of filaments, but other reference directions can arise from factors such as the 3D shape of the parent molecular cloud, the large-scale stratification of the Galactic disk, shock waves driven by nearby supernovae, shearing motions in the ambient medium...
These additional factors will generally tilt filaments away from orientations strictly parallel or perpendicular to the magnetic field.

We can also mention two geometrical reasons.
(1) The 3D orientation of filaments relative to the magnetic field is generally modified upon projection on the sky.
Filaments that are parallel to the magnetic field in 3D remain parallel in projection, but filaments that are perpendicular to the magnetic field in 3D generally do not appear perpendicular in the sky \citep[e.g.,][]{Bracco_2016}.
This projection effect leads to a bias in the histogram of relative orientation.
(2) The filaments observed in 2D intensity or column density maps as well as the magnetic field orientation inferred from 2D polarization maps result from an integration along the line of sight, which generally involves a superposition of several features.
Therefore, relative orientations measured in 2D sky maps are not necessarily representative of the actual situation in 3D space.
The difference between relative orientations in 2D and 3D was recently quantified by \cite{Girichidis_2021} in high-resolution MHD simulations of the turbulent ISM.
Here, the issue of line-of-sight integration may be a real problem for the two distant {\it Herschel} fields, G82 and G202, which lie close to the midplane: in these fields, there is no guarantee that the magnetic field orientation from {\it Planck} truly represents the local magnetic field of the clouds, especially for low $N_{\rm H_2}$ filaments.

\section{Summary and conclusion}
\label{sec:conclusion}

In Paper~1, we presented a new method, called {\tt FilDReaMS}, which makes it possible to detect filaments of a given bar width in an image, to identify the most prevalent bar widths (and corresponding Plummer widths) in the image, and to derive the local orientation angles of the detected filaments.

Here, we applied {\tt FilDReaMS} to a small sample of four {\it Herschel} fields located at different distances (two fields at $\simeq 150\,{\rm pc}$ and two fields beyond $600\,{\rm pc}$), embedded in different Galactic environments, and captured at different evolutionary stages of star formation.
Our fields cover broad dynamic ranges of spatial scales and ${\rm H_2}$ column densities (see Table~\ref{tab:input_fields}), thereby making it possible to detect all kinds of filamentary structures from large and massive filaments down to striations.

In a second step, we compared the filament orientation angles derived with {\tt FilDReaMS} to the local magnetic field orientation angle inferred from {\it Planck} polarization data.
This enabled us to study (for the first time) the statistics of the relative orientation angle as functions of both spatial scale and ${\rm H_2}$ column density.
We emphasize that combining {\it Herschel} and {\it Planck} data despite their widely different angular resolutions does really make sense.
Indeed, the derivation of a filament orientation angle relies on a bar of length larger than $(L_{\rm b})_{\rm min} = 3  \ (W_{\rm b})_{\rm min} = 15\,{\rm px} = 180'' = 3'$, which is nearly half the 7' resolution of {\it Planck}.
This highlights the great potential of combining {\it Herschel} and {\it Planck} data.

No firm, general conclusion can be drawn from our study regarding the preferential widths of filaments.
\begin{itemize}
\item In the nearby fields G210 and G300, no prevalent bar width is found in the range $W_{\rm b} \simeq [0.04, 0.2]\,{\rm pc}$.
\\
\item In the distant fields G82 and G202, two most prevalent bar widths are found in the range $W_{\rm b} \simeq [0.2, 1.1]\,{\rm pc}$: $W_{\rm b}^{\star{\rm peak}} \simeq 0.7\,{\rm pc}$ and $0.9\,{\rm pc}$, corresponding to Plummer widths $2R_{\rm flat} \simeq 1.0\,{\rm pc}$ and $1.3\,{\rm pc}$, respectively, for $p=2.2$. \\
\end{itemize}

Regarding the filament relative orientations to the magnetic field, a few general trends emerge from our study:
\begin{itemize}
    \item Many small (S) filaments are detected in each {\it Herschel} field, over broad ranges of column densities and relative orientations. At low column densities, S filaments tend to be slightly more parallel than perpendicular to the magnetic field. This trend, which increases along the sequence G82, G210, G300, G202, is more pronounced in intensity maps  (which have a twice better angular resolution and can, therefore, reveal smaller filaments) than in column density maps. \\
    \item Only one or two large (L) filaments appear in each {\it Herschel} field, always at high or intermediate column densities. Large filaments clearly show a preferential orientation to the magnetic field, which is nearly perpendicular in G82, G210, and G300, but more nearly parallel in G202. These trends are very similar in column density and intensity maps. \\
    \item The two nearby fields G210 and G300 undergo a transition in relative orientation, from mostly parallel to mostly perpendicular to the magnetic field, at a column density $N_{\rm H_2} \simeq 1.1\times10^{21}\,$cm$^{-2}$ for G210 and $N_{\rm H_2} \simeq 1.4\times10^{21}\,$cm$^{-2}$ for G300. This transition appears more clearly in intensity maps than in column density maps. No such transition is seen in the distant fields G82 and G202.    
\end{itemize}

In the future, we plan to extend our sample to the 116 fields of the {\it Herschel}-GCC program.
This will enable us to probe different places in the Galaxy, with a wide range of physical parameters and across the entire sequence of star formation. This, in turn, will open the way to a complete statistical analysis of the relative orientation between filaments and the local magnetic field, not only as a function of spatial scale, but also as a function of ambient physical conditions and evolutionary stage. In this manner, our results will hopefully contribute to a better physical understanding of the process of filament and star formation.

\begin{acknowledgements}
We extend our deepest thanks to our referee, Gina Panopoulou, for her careful reading of our paper and for her many constructive comments and suggestions.
We also acknowledge useful discussions with Dana Alina, Susan Clark, Mika Juvela, and Julien Montillaud.
The development of Planck has been supported by: ESA; CNES and CNRS/INSU-IN2P3-INP (France); ASI, CNR, and INAF (Italy); NASA and DoE (USA); STFC and UKSA (UK); CSIC, MICINN, JA, and RES (Spain); Tekes, AoF, and CSC (Finland); DLR and MPG (Germany); CSA (Canada); DTU Space (Denmark); SER/SSO (Switzerland); RCN (Norway); SFI (Ireland); FCT/MCTES (Portugal); and PRACE (EU). 
Herschel SPIRE has been developed by a consortium of institutes led by Cardiff University (UK) and including University Lethbridge (Canada); NAOC (China); CEA, LAM (France); IFSI, University Padua (Italy); IAC (Spain); Stockholm Observatory (Sweden); Imperial College London, RAL, UCL-MSSL, UKATC, University Sussex (UK); Caltech, JPL, NHSC, University Colorado (USA). This development has been supported by national funding agencies: CSA (Canada); NAOC (China); CEA, CNES, CNRS (France); ASI (Italy); MCINN (Spain); SNSB (Sweden); STFC (UK); and NASA (USA).
\end{acknowledgements}

\bibliographystyle{aa}
\bibliography{biblio}

\begin{appendix}
\section{Summary of the results obtained from column-density maps}
\label{sec:summary_results_coldensity}

\begin{figure*}
    \centering
    \includegraphics[height=0.95\textheight]{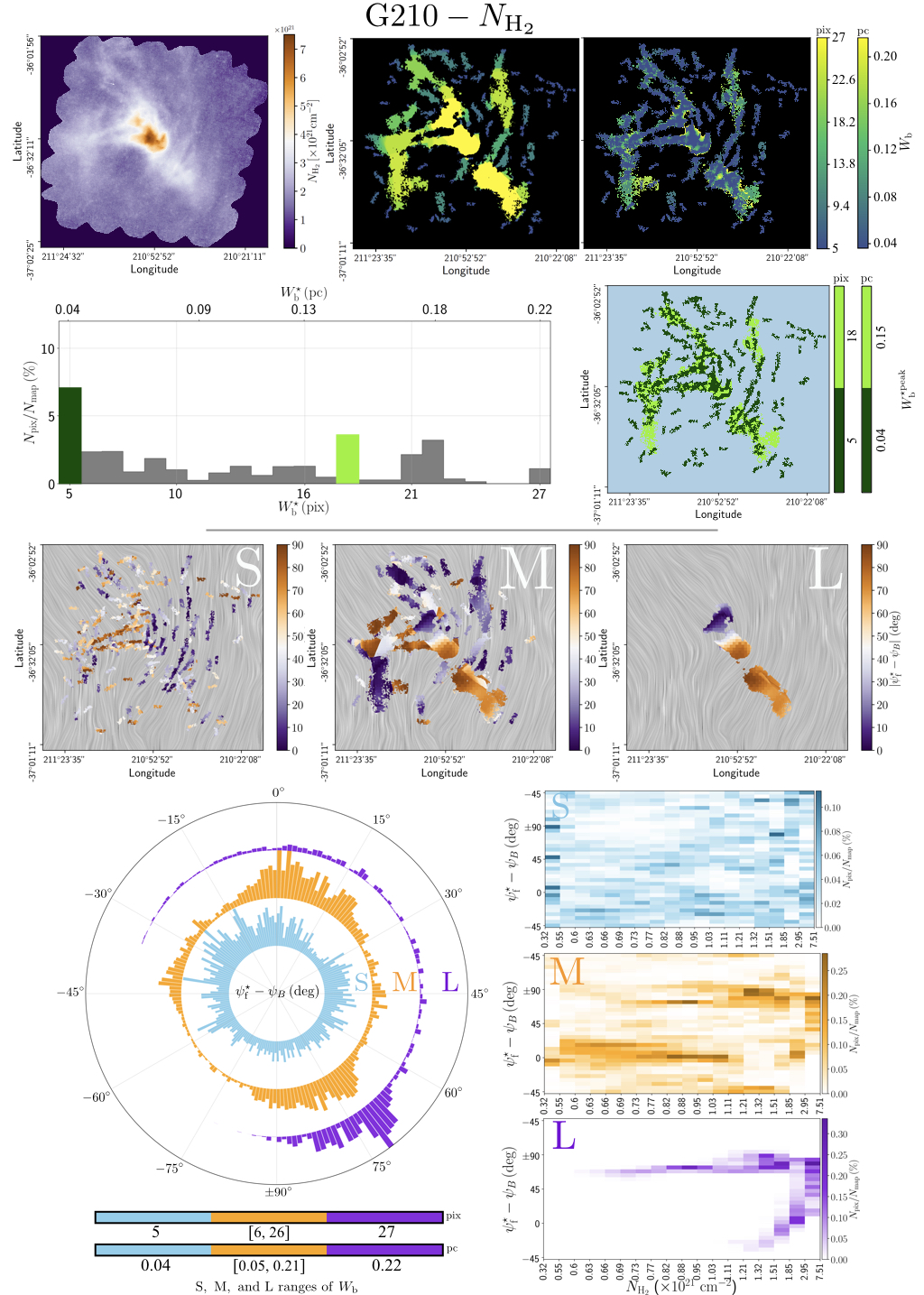}
    \caption{Summary of the main graphical results obtained when applying {\tt FilDReaMS} to the ${\rm H_2}$ column density map of the {\it Herschel} G210 field.
    The top, second, third, bottom-left, and bottom-right panels correspond to Figs.~\ref{fig:filaments_G82}, \ref{fig:filament_best_sizes_G82}, \ref{fig:relative_orientation_map_G82}, \ref{fig:relative_orientation_1D_G82}, and the top row of Fig.~\ref{fig:relative_orientation_2D_G82}, respectively.}
    \label{fig:G210_plots}
\end{figure*}

\begin{figure*}
    \centering
    \includegraphics[height=0.95\textheight]{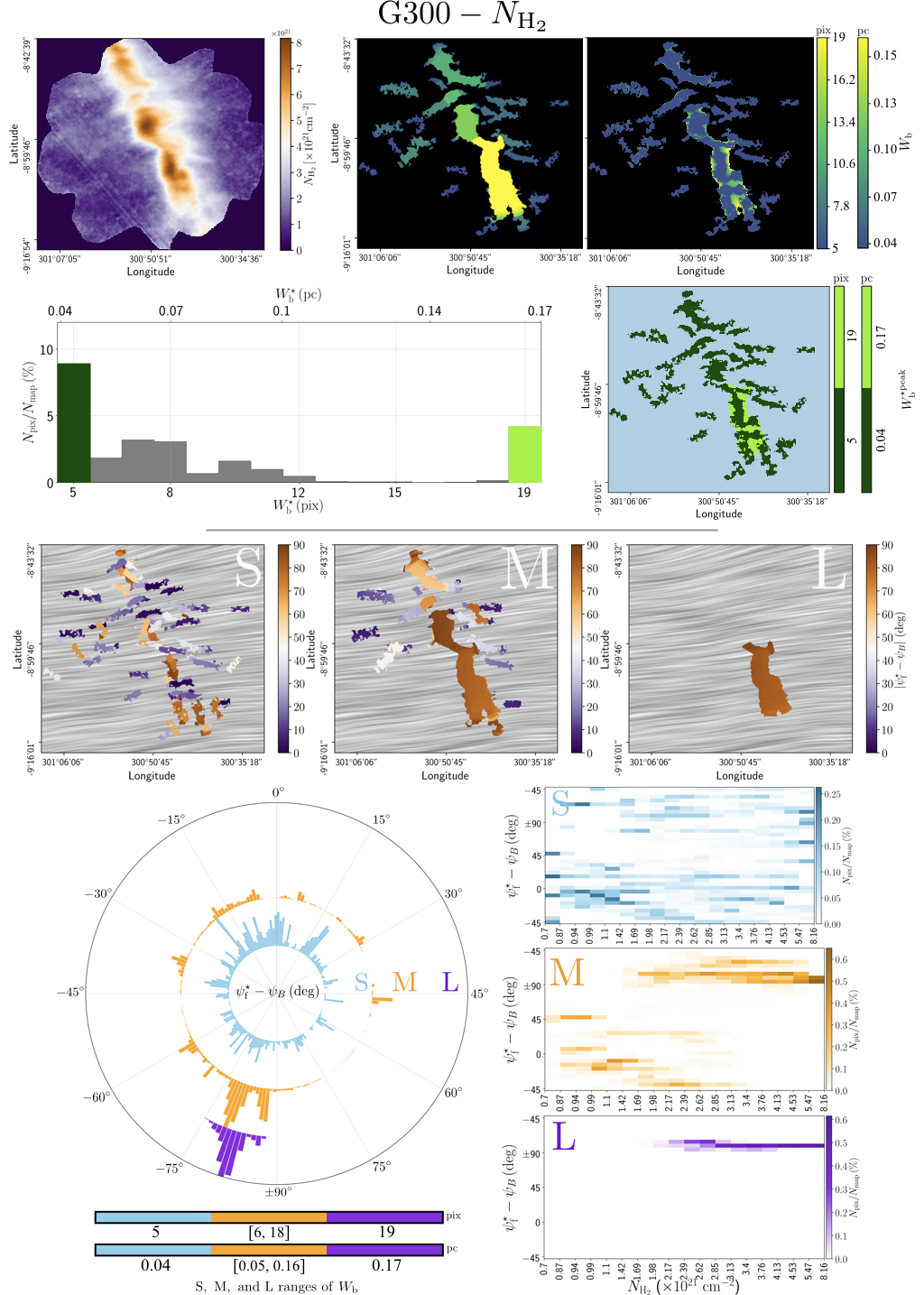}
    \caption{Same as Fig.~\ref{fig:G210_plots}, but for the G300 field.}
    \label{fig:G300_plots}
\end{figure*}

\begin{figure*}
    \centering
    \includegraphics[height=0.95\textheight]{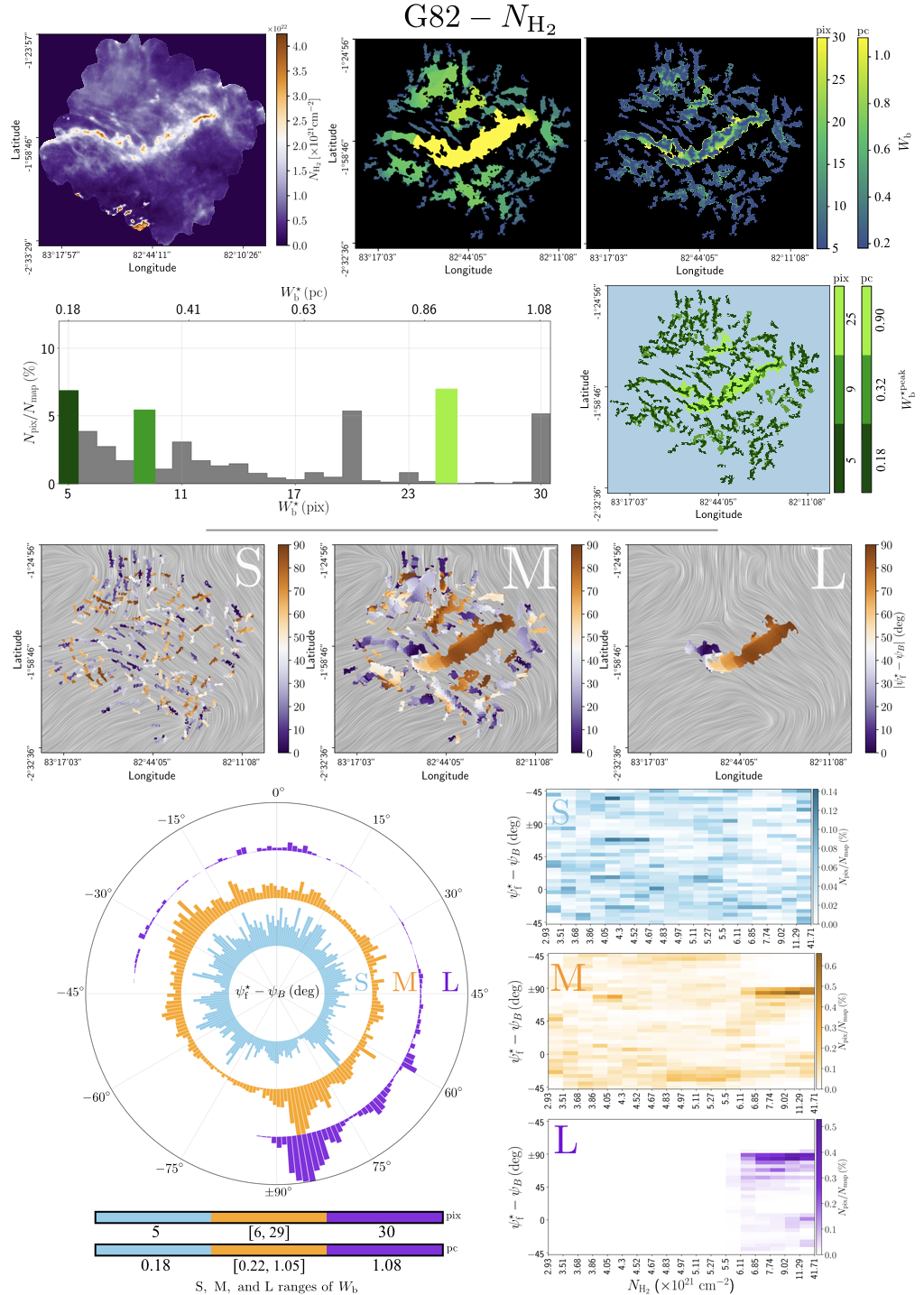}
    \caption{Same as Fig.~\ref{fig:G210_plots}, but for the G82 field.}
    \label{fig:G82_plots}
\end{figure*}

\begin{figure*}
    \centering
    \includegraphics[height=0.95\textheight]{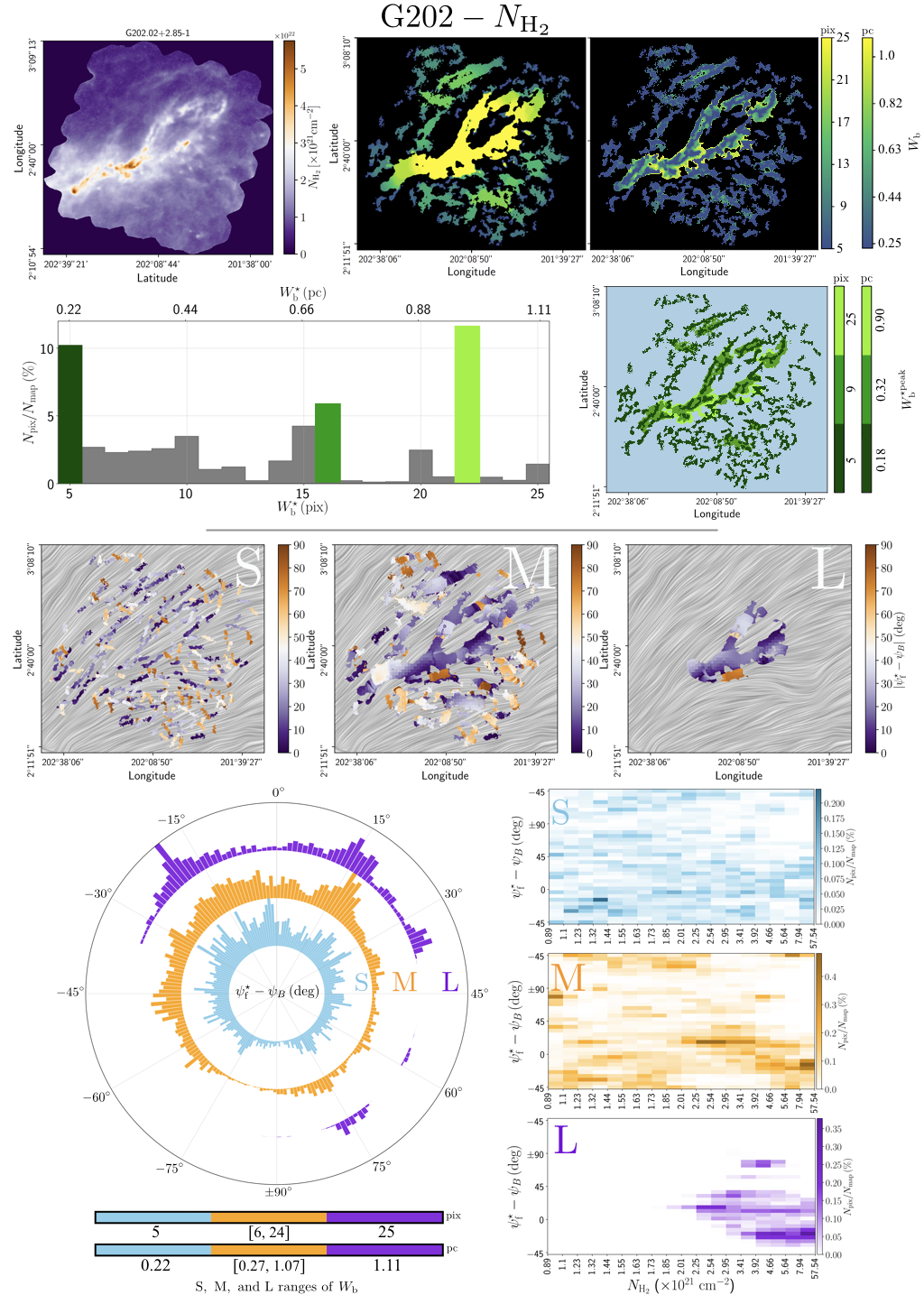}
    \caption{Same as Fig.~\ref{fig:G210_plots}, but for the G202 field.}
    \label{fig:G202_plots}
\end{figure*}

\clearpage
\section{Summary of the results obtained from intensity maps}
\label{sec:summary_results_intensity}
 
\begin{figure*}[ht!]
    \centering
    \includegraphics[height=0.95\textheight]{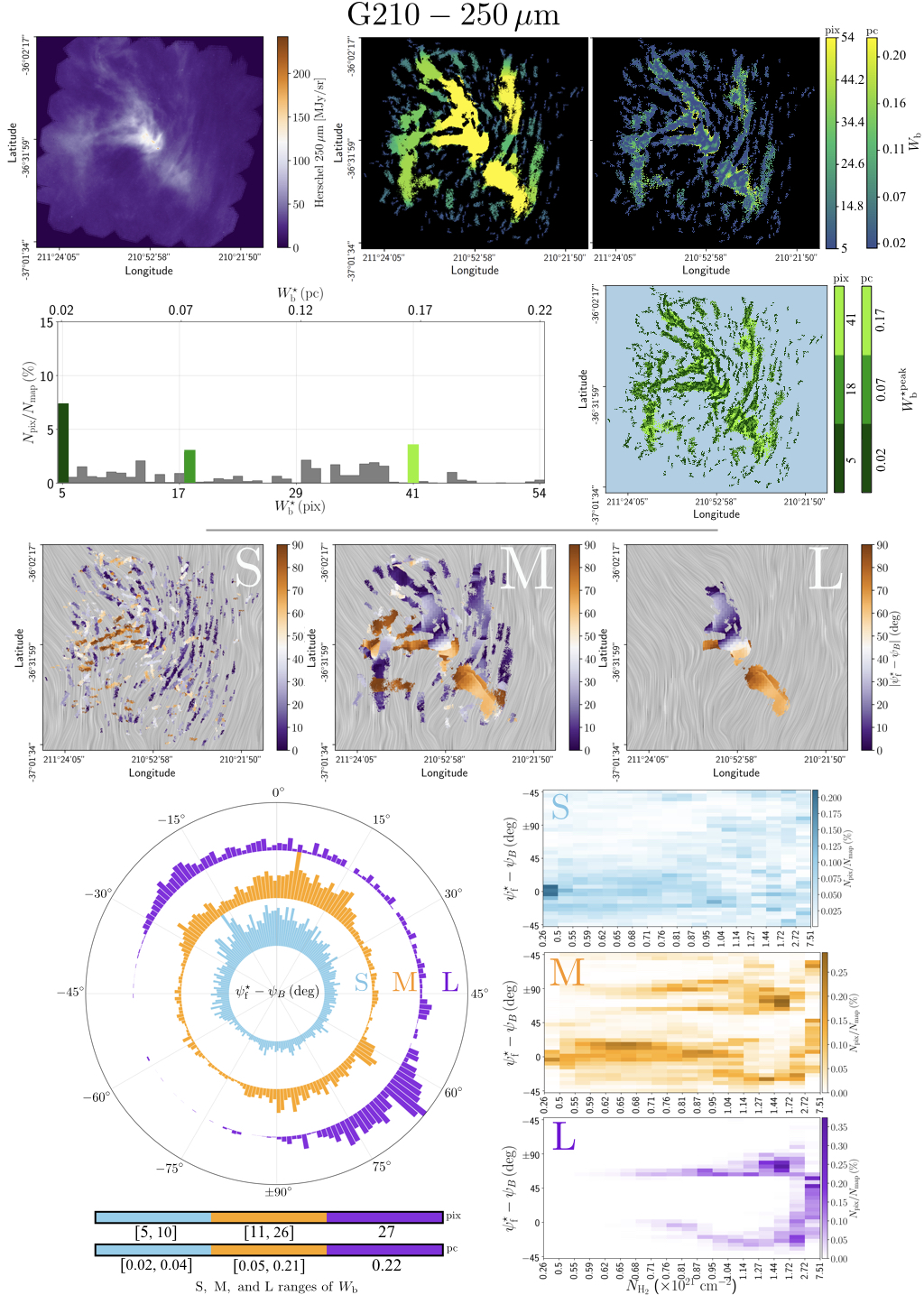}
    \caption{Same as Fig.~\ref{fig:G210_plots}, starting from the {\it Herschel} $250\,{\rm \mu m}$ intensity map of the G210 field.}
    \label{fig:G210_I_plots}
\end{figure*}

\begin{figure*}[ht!]
    \centering
    \includegraphics[height=0.95\textheight]{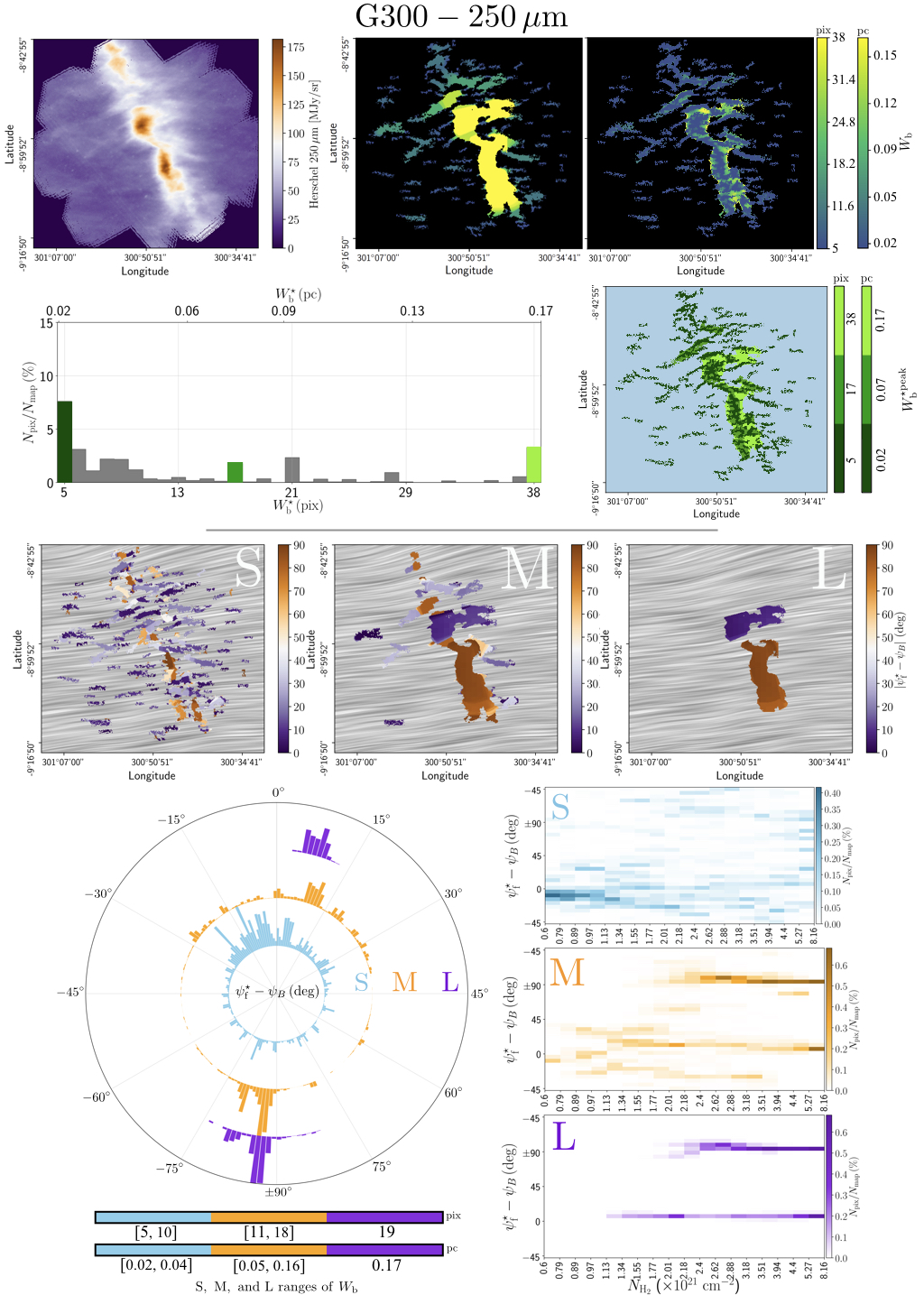}
    \caption{Same as Fig.~\ref{fig:G210_I_plots}, but for the G300 field.}
    \label{fig:G300_I_plots}
\end{figure*}

\begin{figure*}[ht!]
    \centering
    \includegraphics[height=0.95\textheight]{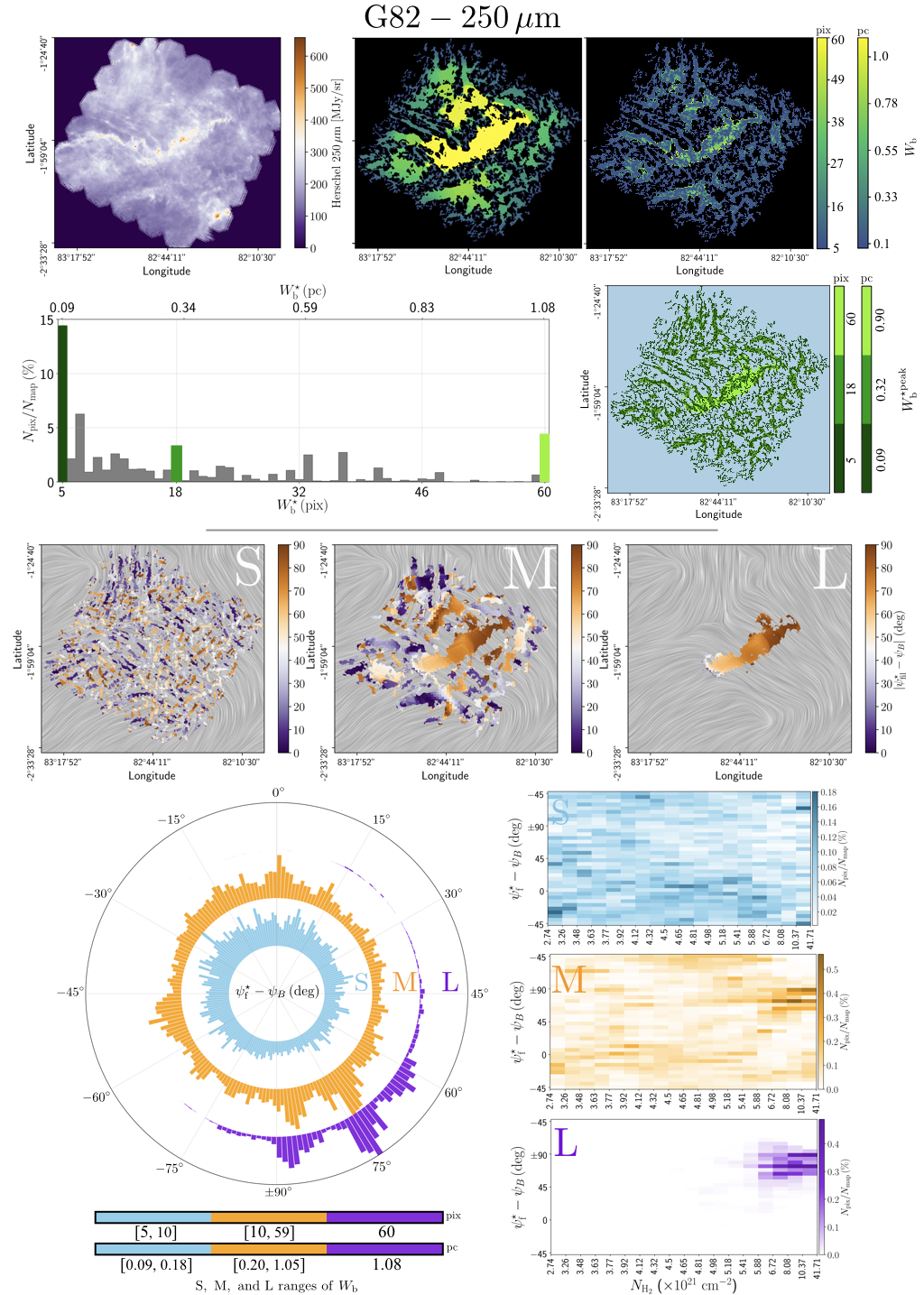}
    \caption{Same as Fig.~\ref{fig:G210_I_plots}, but for the G82 field.}
    \label{fig:G82_I_plots}
\end{figure*}

\begin{figure*}[ht!]
    \centering
    \includegraphics[height=0.95\textheight]{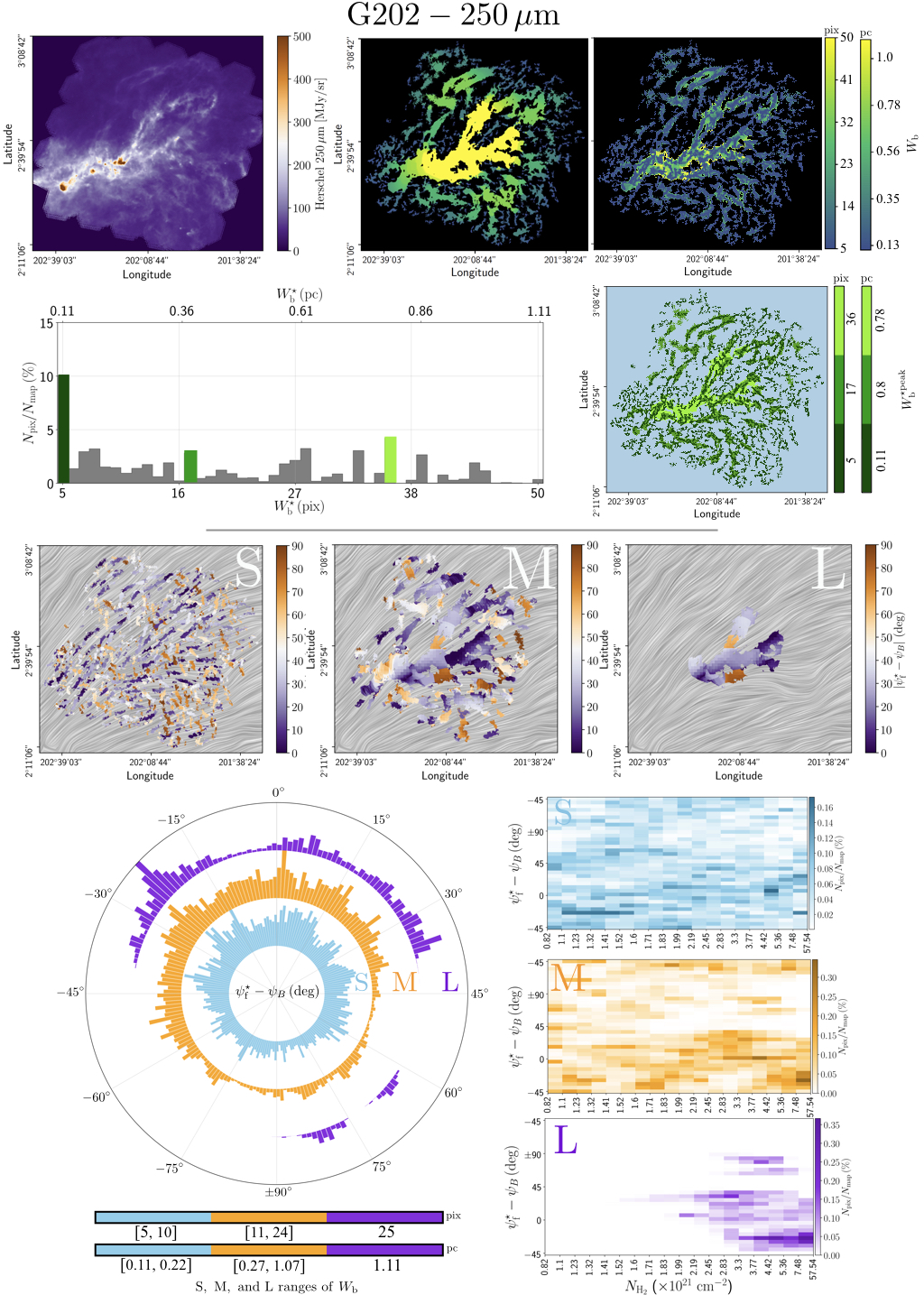}
    \caption{Same as Fig.~\ref{fig:G210_I_plots}, but for the G202 field.}
    \label{fig:G202_I_plots}
\end{figure*}
\end{appendix}

\end{document}